\documentclass[journal]{IEEEtran}

\usepackage{booktabs} % For formal tables
\usepackage{multirow}
\usepackage{algorithm}
\usepackage[noend]{algpseudocode}

\usepackage[noadjust]{cite}

\usepackage{graphicx}
\usepackage[caption=false]{subfig}
\usepackage{wrapfig}
\usepackage{dirtytalk}
\usepackage{tabularx}
\usepackage{amssymb}
\usepackage[usenames, dvipsnames]{color}
\usepackage[export]{adjustbox}

\usepackage{diagbox}

\usepackage[normalem]{ulem}

\usepackage{tikz}
\newcommand\copyrighttext{\scriptsize 
\textcopyright~2020 IEEE. Personal use of this material is permitted. Permission from IEEE must be obtained for all other uses, in any current or future media, including reprinting/republishing this material for advertising or promotional purposes, creating new collective works, for resale or redistribution to servers or lists, or reuse of any copyrighted component of this work in other works. \\
Accepted to be Published in IEEE Transactions on Very Large Scale Integration (VLSI) Systems, DOI: 10.1109/TVLSI.2020.2968552}
\newcommand\copyrightnotice{%
\begin{tikzpicture}[remember picture,overlay]
\node[anchor=south,yshift=0pt] at (current page.south) {\fbox{\parbox{\dimexpr\textwidth-\fboxsep-\fboxrule\relax}{\copyrighttext}}};
\end{tikzpicture}%
}

\begin{document}
\title{SAT-hard Cyclic Logic Obfuscation for Protecting the IP in the Manufacturing Supply Chain \\}

\author{
    \IEEEauthorblockN{
        Shervin~Roshanisefat\IEEEauthorrefmark{1}, 
        Hadi~Mardani~Kamali\IEEEauthorrefmark{1}, 
        Houman Homayoun\IEEEauthorrefmark{2}, 
        and~Avesta~Sasan\IEEEauthorrefmark{1}}
    \IEEEauthorblockA{
            \begin{center}
            \begin{tabular}{ c c }
                \IEEEauthorrefmark{1}Department of Electrical and Computer Engineering &
                \IEEEauthorrefmark{2}Department of Electrical and Computer Engineering \\
                George Mason University, Fairfax, VA, U.S.A. & 
                University of California, Davis, CA, U.S.A.\\
                e-mail: \{sroshani, hmardani, asasan\}@gmu.edu &
                e-mail: hhomayoun@ucdavis.edu \\
            \end{tabular}
            \end{center}
    } 
}

\IEEEaftertitletext{\vspace{-2\baselineskip}}

% The paper headers
% \markboth{IEEE TRANSACTIONS ON VERY LARGE SCALE INTEGRATION (VLSI) SYSTEMS,~Vol.~-, No.~-, August~-}%
% {Roshanisefat \MakeLowercase{\textit{et al.}}: IEEE TRANSACTIONS ON VERY LARGE SCALE INTEGRATION (VLSI) SYSTEMS}

\maketitle

\begin{abstract}
State-of-the-art attacks against cyclic logic obfuscation use satisfiability solvers that are equipped with a set of cycle avoidance clauses. These cycle avoidance clauses are generated in a pre-processing step and define various key combinations that could open or close cycles without making the circuit oscillating or stateful. In this paper, we show that this pre-processing step has to generate cycle avoidance conditions on all cycles in a netlist, otherwise, a missing cycle could trap the solver in an infinite loop or make it exit with an incorrect key. Then, we propose several techniques by which the number of cycles is exponentially increased as a function of the number of inserted feedbacks. We further illustrate that when the number of feedbacks is increased, the pre-processing step of the attack faces an exponential increase in complexity and runtime, preventing the correct composition of cycle avoidance clauses in a reasonable time. On the other hand, if the pre-processing is not concluded, the attack formulated by the satisfiability solver will either get stuck or exit with an incorrect key. Hence, when the cyclic obfuscation under the conditions proposed in this paper is implemented, it would impose an exponentially difficult problem for the satisfiability solver based attacks.
\end{abstract}


\IEEEpeerreviewmaketitle

\copyrightnotice

\section{Introduction}
The cost of building a new semiconductor fab was estimated to be \$5.0 billion in 2015, with large recurring maintenance costs \cite{DIGITIMES,6926108}, and sharply increases as technology migrates to smaller nodes. Thus, to reduce the fabrication cost, and for economic feasibility, most of the manufacturing and fabrication is pushed offshore \cite{DIGITIMES}. However, many offshore fabrication facilities are considered to be untrusted, which has raised concern over potential attacks in the manufacturing supply chain, with an intimate knowledge of the fabrication process, the ability to modify and expand the design before production, and unavoidable access to the fabricated chips during testing. Hence, fabrication in untrusted fabs has introduced multiple forms of security threats from supply chain including that of overproduction, Trojan insertion, reverse engineering, intellectual property (IP) theft, and counterfeiting \cite{6926108}.

To prevent the adversaries from such attacks, researchers have proposed various obfuscation methods for hiding and/or locking the functionality of a netlist \cite{threatszamiri,8203496,Keshavarz2018,srcRoshanisefat,comakimia}. However, the validity and strength of logic obfuscation to defend an IP against adversaries in the manufacturing supply chain was seriously challenged as researchers demonstrated that the de-obfuscation attacks leveraging satisfiability (SAT) solvers \cite{7140252,el2015integrated,8474189} combined with Signal Probability Skew (SPS) attacks~\cite{7858346} could break the existing obfuscation schemes (both locking and camouflaging) in a relatively short time. Cyclic obfuscation \cite{Shamsi:2017:COC:3060403.3060458} was another approach that was considered as a defense mechanism against SAT solvers. However, this technique was later broken by CycSAT attack~\cite{8203759}. CycSAT added a pre-processing step to the original SAT attack for detection and avoidance of cycles in the netlist before deploying an SAT attack. In this paper, we illustrate that the pre-processing step of CycSAT attack has to process a cycle avoidance condition for every cycle in the netlist, otherwise, the subsequent SAT attack could get stuck in an infinite loop or returns UNSAT. Hence, the runtime of the pre-processing step is linearly related to the number of cycles in a  netlist. Besides, we illustrate that the generation of a cycle avoidance clause for a netlist of cyclic Boolean nature is far more time consuming than an acyclic Boolean logic.

From this observation, we first propose several mechanisms for cyclification of a non-cyclic Boolean netlist. Then, we propose two design techniques by which a linear increase in the number of inserted feedbacks in a netlist would exponentially increase the number of generated cycles. Since a successful SAT attack on a cyclic circuit requires the generation of a per-cycle avoidance clause and considering that our proposed techniques make the time it takes to generate such avoidance clauses an exponential function of the number of inserted feedbacks, CycSAT attack faces exponential runtime at its processing step. Hence, when deploying CycSAT, the complexity of the pre-processing of the resulting cyclic netlist goes beyond a reasonable time limit. On the other hand, skipping the prepossessing result in an unsuccessful SAT attack. Hence, cyclic obfuscation, when constructed using the proposed methodology, proves to be a strong defense against the SAT and CycSAT attack. 

Contributions of this paper are as follows: 1) We provide a comprehensive background on (both cyclic and acyclic) SAT-resistant logic-encryption solutions, 2) we introduce a new attack that enables a SAT attack to break two recently published logic locking solutions (i.e., obfuscation using nested or hard cycles), 3) we propose a new cyclic obfuscation solutions that make the number of created (real and dummy) cycles an exponential function of the number of inserted feedbacks, and elaborate how it is as an effective mean for breaking cyclic SAT attacks, 4) we propose a timing-aware cyclification algorithm to manage and control the timing overheads of our proposed solution, 5) we assess the effectiveness of our proposed solutions on several benchmarks using an improved/modified cyclic attack and report the power, performance and area overhead of our proposed solution.

The rest of this paper is organized as follows. In section \ref{background}, we cover the background on logic obfuscation. Then, in section \ref{breakingCycSAt}, we elaborate on the limitation of cyclic attacks and our approach for breaking/preventing these attacks. In section \ref{cyclock}, we introduce our techniques for building an exponential relation between the number of feedbacks and the number of created cycles in a circuit. We also introduce three mechanisms for building a cyclic Boolean function to further increase the complexity of pre-processing in cyclic attacks. Our experimental results are summarized in section \ref{results}. Section \ref{conclusion} concludes the paper.
\vspace{-2mm}
\section{Background} \label{background}
Logic obfuscation is the process of hiding the functionality of an IP by building ambiguity or by implementing post-manufacturing means of control and programmability into a netlist. Gate camouflaging \cite{rajendran2013security,6881480,7495587,li2017provably} and circuit locking \cite{Yasin_sfll,8429401} are two of the widely explored obfuscation mechanisms for this purpose. A camouflaged gate is a gate that after reverse engineering (using delayering and lithography) could be mapped to any member of a possible set of gates or may look like one logic gate (e.g., AND), however functionally perform as another (e.g., XOR). In locking solutions, the functionality of a circuit is locked using several key inputs such that only when a correct key is applied, the circuit resumes its expected functionality. Otherwise, the correct function is hidden among many of the $2^K$ ($K$ being the number of keys) circuit possibilities. The claim raised by such an obfuscation scheme was that to break the obfuscation, an adversary needs to try a large number of inputs and key combinations to extract the correct key, and the difficulty of this process increases exponentially as the number of keys and primary inputs increases. Hence, if enough gates are obfuscated, an adversary faces an unacceptably long time (claimed as years to decades) to break the obfuscation scheme. Note that the availability of scan chains, which is inserted following Design for Test (DFT) recommended flow, allows an adversary to access combinational logic in each stage of a sequential circuit, load the desired input, execute the stage for one cycle, and readout the output.

The validity and strength of logic obfuscation to defend the IP against adversaries in the manufacturing supply chain was seriously challenged as researchers demonstrated that the SAT solvers, when formulated according to Algorithm \ref{SAT_algoritm}, could break the obfuscation (both locking and camouflaging) in a matter of minutes as opposed to the promised claim of years and decades \cite{7140252,el2015integrated}. In this algorithm, $C(X,K,Y)$ refers to the obfuscated circuit that produces output vector $Y$ using input vector $X$ and key vector $K$ and $C_{BlackBox}(X)$ refers to the output of the activated circuit for input vector $X$. As illustrated in Algorithm \ref{SAT_algoritm}, to employ a SAT attack, the obfuscated circuit is transformed into a circuit SAT problem, in which the SAT solver looks for an input value X for which the obfuscated circuit produces two different outputs for two different input keys. Such input is referred to as a \emph{Distinguishing Input} (DIP) $X_{DI}$. Each time a new $X_{DI}$ is found, the circuit SAT is updated to make sure that the next two keys that will be found in the next iteration of SAT solver invocation, produce the same output for all previously discovered $X_{DI}$. This is done by building a Distinguishing Input Validation Circuit (DIVC) as illustrated in Algorithm \ref{SAT_algoritm}. When the SAT solver can no longer find a $X_{DI}$, the DIVC circuit contains a complete set of distinguishing inputs. At this point, any key that satisfies the DIVC (by calling a SAT solver on this circuit) is the key to the obfuscated circuit \cite{7140252,el2015integrated,smtazar}.

\begin{algorithm}[t]
\caption{SAT Attack on Obfuscated Circuits \label{SAT_algoritm}}
\begin{algorithmic}[1]
\scriptsize
\State $DIVC = 1$;
\State $SAT_{circuit} = C(X,K_1,Y_1) \wedge C(X,K_2,Y_2) \wedge (Y_1 \ne Y_2)$;
\While {$((X_{DI},K_1,K_2)\leftarrow SAT_F(SAT_{circuit})=T)$} 
    \State $Y_f \leftarrow C_{BlackBox}(X_{DI})$; 
    \State $DIVC = DIVC \wedge C(X_{DI},K_1,Y_f) \wedge C(X_{DI},K_2,Y_f)$;
    \State $SAT_{circuit} = SAT_{circuit} \wedge DIVC$;
\EndWhile

\State $KeyGenCircuit = DIVC \wedge (K_1 = K_2)$
\State $Key \leftarrow SAT_F (KeyGenCircuit)$
\end{algorithmic}
\end{algorithm}

\vspace{-13pt}
\subsection{Acyclic Logic Obfuscation}
The revelation of this attack redirected the attention of the researchers to find harder obfuscation schemes that protect acyclic Boolean logic and resist the SAT attack. These methods have targeted a number of weaknesses in the SAT attack and could be categorized into three categories: \\[-7pt]

\textbf{1- Weaker Distinguishing Inputs:} Original SAT attack was powerful because each DIP could rule out several wrong keys and constrain the key space effectively. The SARLock and Anti-SAT \cite{7495588,xie2016mitigating} logic locking methods were proposed to mitigate this vulnerability. In a circuit protected by these solutions, a wrong key produces a wrong output only for one input. This will create a much weaker DIP as each DIP can only rule out one wrong key. Hence, a SAT attack will be reduced to a Brute-force attack as it requires an exponential number of DIPs to find the correct key. A design protected by these mechanisms, regardless of the key used for its activation, behaves very similar to the original design (except for one input). Hence, this group of obfuscation solutions suffers from low output corruption. To increase the output corruption, they could be augmented with other (output corruption oriented) obfuscation mechanisms. However, by using approximate SAT attack \cite{7951805} almost all key values for the augmented obfuscation mechanism could be correctly identified.

Further research revealed that these obfuscation techniques are vulnerable to removal \cite{7858346}, Bypass \cite{bypass} and FALL \cite{fall_attack} attacks. In a removal attack, these SAT hard blocks are identified using Signal Probability Skew (SPS) attack \cite{7858346} and removed. In Bypass attack \cite{bypass}, an auxiliary circuit that recovers the wrong output in these locking schemes is created. This attack identifies the input combinations that produce the wrong output for a wrong key; then it adds a bypass circuit to flip the wrong output when that specific input is applied. In FALL attacks, a functional analysis of the circuit will be performed and have two stages. In the first stage, it analyses the functionality of the obfuscated circuit and tries to identify the locking keys. If there was more than one candidate for the locking key, it tries to use the SAT to find the correct locking key from a list of alternatives and using simulations on the unlocked circuit.

% \textcolor{red}{The main issue with these locking schemes is their low output corruptibility to prevent DIs from pruning the key space during the SAT attack. By using this leverage to create a SAT resilient circuit, the locked circuit will be more likely vulnerable to removal or Bypass attacks. Hence, there is a trade-off between the protected patterns and the security level of the locked circuit.}

\textbf{2- Increasing Circuit-SAT Complexity:}
Another feature that makes the SAT attack powerful is the fast execution time of the underlying SAT solver in solving the circuit SAT and extracting DIPs. For locking schemes in this category, the netlist is designed in a way that translates to a large circuit-SAT with possibly a SAT-hard portion and thus requires more time to solve. Cross-Lock \cite{crosslock} exploited this vulnerability by adding cross-bars to the netlist and obfuscates circuit connections. Equivalent circuit-SAT in this method requires large multiplexers and the symmetric nature of this block will make it a SAT-hard problem \cite{Kamali:2019:FHD:3316781.3317831,kolhe2019security,kolhecustom}. Without any additional clauses, any SAT solver requires a long execution time to find a single distinguishing input.

Netlists with camouflaged or memory-based blocks could also be used for this purpose. For these blocks, an equivalent circuit should be used that replaces them. For blocks with a large number of input and key ports, the equivalent circuit could be very large. This is especially true in the case of a locked circuit with large LUTs. This could lead to a large circuit-SAT with lots of SAT clauses.\\[-7pt]

\textbf{3- SAT Unsolvable Structures:}
SAT attack needs to translate the reverse-engineered netlist into CNF clauses to be able to use the underlying SAT solver. Memory-blocks and Boolean gates could be easily translated into CNF clauses using equivalent circuits and Tseitin \cite{Tseitin1983} transformation. Boolean limitation of SAT solvers could be used as a vulnerability to implement non-Boolean structures to counter the SAT attack.

Delay Locking \cite{delaylocking} is one of such methods. It uses key-gates to lock both the functionality and the timing behavior of the obfuscated circuits. The logic aspect of the locking could be easily translated to CNF, however, the behavioral (timing) aspect of circuit operation can not be easily translated into a SAT friendly CNF. Hence, formulating a SAT attack on a delay-locked netlist will produce a circuit of correct functionality, but the timing violations will make the circuit malfunctioning. This method could potentially prevent overproduction or any reuse of fabrication materials like masks, but it can not prevent reverse engineering and IP-theft of the design. Also, an attack called TimingSAT \cite{TimingSAT} was later proposed to break this obfuscation method.

\subsection{Cyclic Logic Obfuscation} \label{cycobfuscation}
Another method that could render SAT solvers ineffective is to invalidate the acyclic nature of netlist by using cyclic logic obfuscation. Cyclic logic obfuscation was first proposed in \cite{Shamsi:2017:COC:3060403.3060458} whereby introducing feedbacks in the netlist, the netlist is no longer a Directed Acyclic Graph (DAG). In their approach, each intentionally created cycle had more than one way to be opened, making such cycle irreducible by structural analysis, claiming that the existence of such a cycle breaks the original SAT attack in \cite{7140252,el2015integrated}. 

Attacks previously proposed for breaking logic locking solutions are not effective on cyclically obfuscated circuits. The brute-force attack on obfuscated circuits (even those that are not SAT-hard) will face exponential difficulty. The sensitization attack would not work on cyclic circuits since key values control the multiplexers' select line and the select values can not be sensitized to output pins. The pure SAT attack does not work on cyclic circuits as cycles could either trap the SAT solver or make it exit with an incorrect key, a problem that also occurs in approximate SAT attacks (i.e., AppSAT); the approximate attacks address the issue of separating the keys between SAT-hard and conventional obfuscation. Considering that cyclic circuits trap the SAT solver, this group of attack is also would not work. Removal and SPS attacks are aimed at detecting and removing point functions which are used as a means of building SAT hard solutions in the DAG-based network. Considering that the cyclic obfuscation does not use a point function, SPS and removal attacks are not applicable.

Cyclic obfuscation was later broken with introduction of cyclic (cycle-aware) attacks in \cite{Chen:2018:ESA:3217208.3190853,besat,8203759}. CycSAT was the first cyclic attack, details of which are shortly discussed. Later, Chen \cite{Chen:2018:ESA:3217208.3190853} introduced an enhanced SAT-attack that considers structural cycles. From a functional standpoint, this attack acts similar to the structural attack in CycSAT. 

\begin{figure*}
    \centering
    \subfloat[]{\includegraphics[width=0.25\columnwidth]{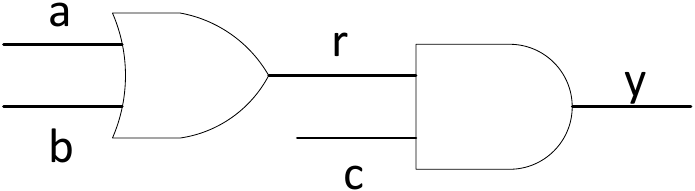}}
    \hspace{0.2cm}
    \subfloat[]{\includegraphics[width=0.55\columnwidth]{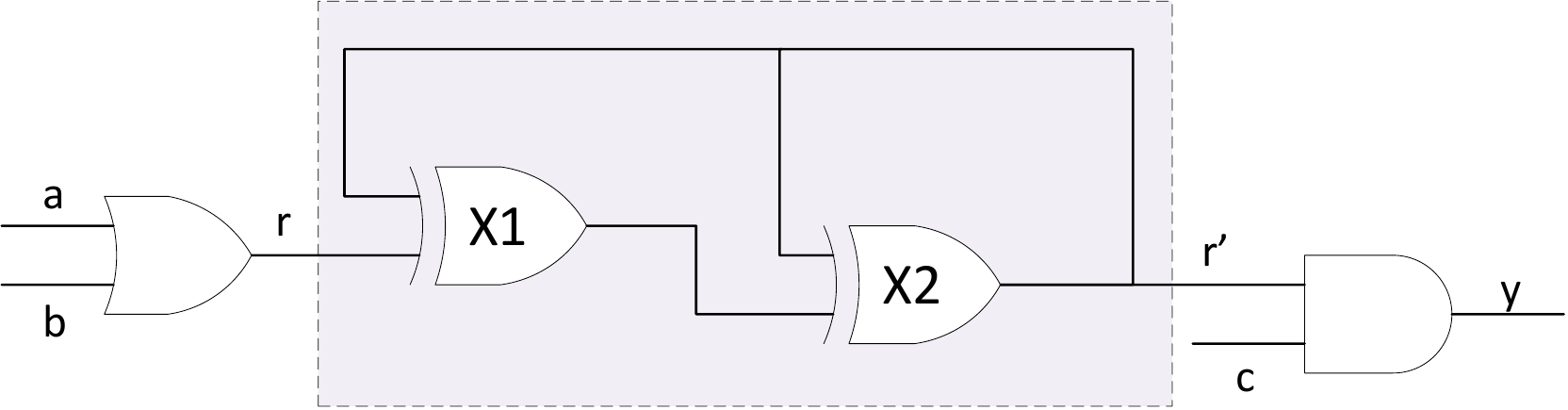}}
    \hspace{0.2cm}
    \subfloat[]{\includegraphics[width=0.35\columnwidth]{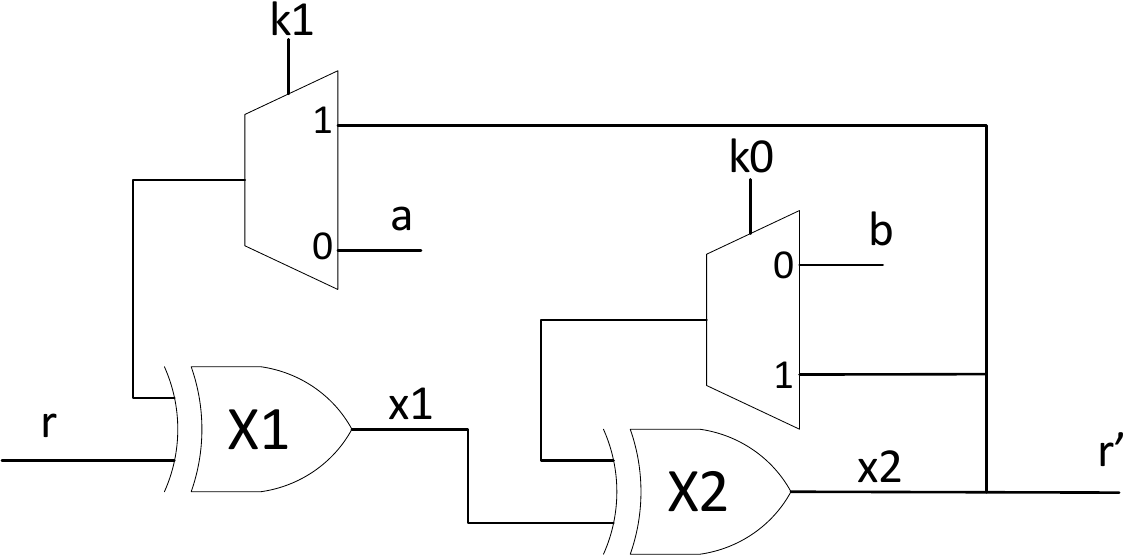}}
    \hspace{0.2cm}
    \subfloat[]{\includegraphics[width=0.35\columnwidth]{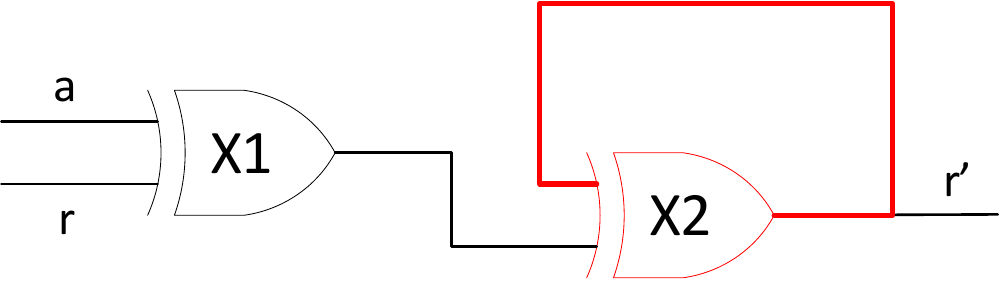}}
    \hspace{0.2cm}
    \subfloat[]{\includegraphics[width=0.35\columnwidth]{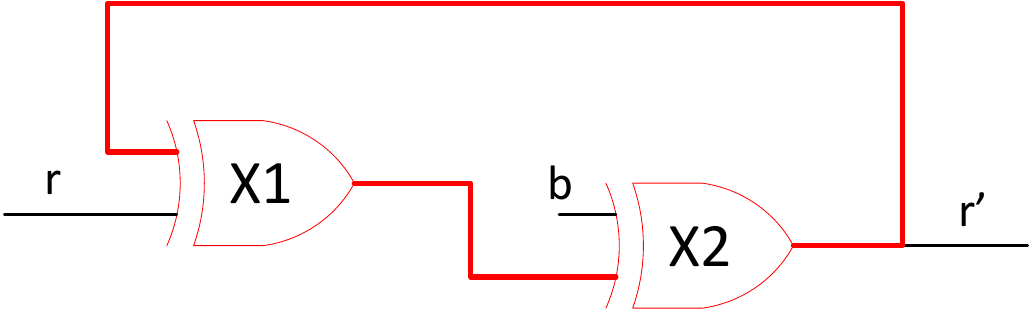}}
    \caption{Cyclification with dependent cycles: (a) Original circuit, (b) cyclified with an auxiliary-circuit that acts as a buffer, (c) Obfuscated auxiliary circuit, (d) auxiliary circuit with broken outer cycle, (e) auxiliary circuit with broken inner cycle. \vspace{4mm}}
    \label{dependcycle_break}
\end{figure*}

In CycSAT attack, before invoking the SAT solver, the netlist is checked for key conditions that may result in the creation of cycles. These conditions are translated to a set of cycle avoidance clauses and are added to the list of clauses that represent the circuit SAT problem. The algorithm \ref{CycSAT_algoritm} illustrates the flow of utilizing the cycle avoidance-clauses in CycSAT.

\begin{algorithm}[t]
\caption{CycSAT Attack on Cyclic Obfuscated Circuits \label{CycSAT_algoritm}}
\begin{algorithmic}[1]
\scriptsize
\State Find a set of feedback signals $(w_0, w_1, ...w_m)$;
\State Compute "no structural path" formulas $F(w_0, w'_0)$, ..., $F(w_m, w'_m)$;
\State $NC(K)=\wedge^m_{i=0}F(w_i,w'_i)$
\State $C(X,K,Y) = C(X,K,Y)\wedge NC(K)$
\State $SAT_{circuit} = C(X,K_1,Y_1) \wedge C(X,K_2,Y_2) \wedge (Y_1 \ne Y_2)$;
\While {$((X_{DI},K_1,K_2)\leftarrow SAT_F(SAT_{circuit})=T)$} 
    \State $Y_f \leftarrow C_{BlackBox}(X_{DI})$; 
    \State $DIVC = DIVC \wedge C(X_{DI},K_1,Y_f) \wedge C(X_{DI},K_2,Y_f)$;
    \State $SAT_{circuit} = SAT_{circuit} \wedge DIVC$;
\EndWhile

\State $KeyGenCircuit = DIVC \wedge (K_1 = K_2)$
\State $Key \leftarrow SAT_F (KeyGenCircuit)$

\end{algorithmic}
\end{algorithm}

In this algorithm, $(w_0, w_1, ...w_m)$ is a collection of feedback signals whose break will make the encrypted circuit acyclic and $w'_i$ is a signal that feeds to $w_i$ before the break. The function $F(w_i,j)$ is a function that construct the condition for "\emph{having no structural path}" between signal $w_i$ to signal $j$. The  $F(w_i,j)$ is computed by starting from a feedback signal $w_i$ and constructs a string of clauses that satisfy the following condition while traversing a cycle:

\begin{equation} \label{eq:erl}
\footnotesize
F(w_i,j) = \bigwedge_{l\in NK(j)}F(w_i,l)\lor bk(l,j)
\end{equation}

In this function, the $NK(j)$ are the non-key inputs of signal j, and $bk(l,j)$ is the condition on the key assuring key does not affect j. This function is initiated with condition $F(w_i,w_i)=0$ and finishes after completing the loop. In this case, the condition for no structural path is tested on all discovered feedback signals in line 3 of the algorithm.

Subsequently, Rezaie et al. proposed two solutions \cite{ZhouCycles,CycSATunresolvable} to counter CycSAT attack. In the first solution \cite{ZhouCycles}, by adding hard cycles to the original netlist, they create a situation that any traversal of the feedback signals will miss a cycle. Also, for this method, dependent cycles are added to the original circuit such that two nested cycles should be closed to create a working circuit. In the second solution \cite{CycSATunresolvable}, a method is introduced to create cycles that behave non-combinational in unreachable states. However, in the next section, after providing further detail on these locking mechanisms, we illustrate that these solutions are still vulnerable and a simple modification to CycSAT attack could easily break them.

Finding all cycles in a cyclic circuit (a requirement for CycSAT attack) is not an easy task. Recently, Shen et al. introduced a new attack called BeSAT~\cite{besat}. Authors of this attack argue that \say{it is impossible to capture all cycles in any graph with any set of feedback signals as done in CycSAT algorithm}.  To address this problem, BeSAT first adds \say{no structural path} (CycSAT-I) conditions for a \say{set of feedback signals}. This is similar to the pre-processing step in CycSAT attack. Then, it performs SAT while monitoring the behavior of the attack: during the DIP generation process, due to the missing \textit{NC} clauses, it is possible that solving the circuit-SAT problem results in repeated DIPs. Under the original SAT attack, this could trap the attack in an infinite loop. In BeSAT, every new DIP is compared with previous DIPs and if it was generated before, the algorithm uses it to determine the stateful key $K_s$. BeSAT compares the output of the new DIP for the two found keys with the oracle circuit. The output of the stateful key disagrees with the oracle circuit. Then, the found stateful key will be explicitly banned by adding ($K1 \neq K_s~\wedge~K2 \neq K_s$) condition to the circuit-SAT problem. After finding all DIPs and banning all stateful keys, BeSAT begins pruning oscillating keys by employing ternary SAT.

%To break CycSAT, in this paper, we focus our attention on two fronts (1) the pre-analysis step is now a part of SAT solution, hence if the execution time of pre-analysis step, its memory requirements, or the number of generated avoidance-clauses could be increased exponentially with the number of loops in the circuit, we are still capable of building a SAT hard solution. (2) CycSAT relies on the assumption that the original circuit is a DAG. However, as shown in  \cite{Cyclicbooleancircuits}\cite{Rivest:1977:NFM:1310165.1310794}\cite{7406959}\cite{1466160}\cite{1218927} a Boolean circuit could have a cyclic topology. In this case, by introducing both functionally-needed and structurally-obfuscating cycles, the pre-analysis step that relies on loop free DAG nature of the original circuit is invalidated. This results in producing incorrect avoidance-clauses for the subsequent invocation of a SAT solver.

% \vspace{-10pt}
\section{Analyzing the Weaknesses of Cyclic Obfuscation} \label{breakingCycSAt}
In this section, we first show that the nested cycles could not guarantee a secure cyclic obfuscation. Furthermore, we propose a new attack mechanism to break the hard cycles. Then, we investigate the weaknesses of CycSAT attack, according to which we propose a new mechanism for cyclic obfuscation.

\vspace{-6pt}
\subsection{Breaking Nested Cycles}\label{nested_cycles}
An obfuscation method that was previously proposed to counter CycSAT attack is the use of nested cycles \cite{ZhouCycles}. In this method, the original circuit is augmented with a pair of nested cycles such that for correct operation, both cycles should be closed. An example of such a transformation is shown in Fig. \ref{dependcycle_break}.b for the original circuit in Fig. \ref{dependcycle_break}.a. After the transformation, the nested cycles are a needed and valid part of the original circuit and attempting to remove one or both cycles will affect the correct functionality of the circuit. A designer may try to obfuscate these cycles using multiplexers as shown in Fig. \ref{dependcycle_break}.c.

Direct application of structural CycSAT attack, as claimed by authors in \cite{ZhouCycles} results in breaking each of nested cycles separately, creating an oscillating and un-SAT-isfiable circuit. However, as claimed earlier, we can still deploy a successful attack against this variant of cyclic obfuscation using a simple modification to the pre-processing step of CycSAT attack.

\begin{algorithm}[t]
\caption{Generating RC Clauses for Dependent Cycles \label{nested_alg}}
\begin{algorithmic}[1]
\scriptsize
    \Procedure{reduction\_attack}{circuit $K$}
    \State Find and sort all cycles in $K$ by their length $C=(c_0, c_1, ...c_m)$;
    \ForAll {$c_i$ in $C$}
        \State $RC(c_i)=\phi$;
    \EndFor
    \ForAll {$c_i$ in $C$}
        % \State $NC(c_i)=\phi$
        \If {IS\_COMB\_CYCLE($c_i$) == False};
            \State $RC(c_i)=RC(c_i)~\lor~opened(c_i)$;
            \While{$c_j \leftarrow$ next outer cycle}
                \State $sub\_circuit \leftarrow$ sub-circuit of closed $c_i$ and $c_j$;
                \If{IS\_COMB\_CYCLE($sub\_circuit$)}
                    \State $RC(c_i)=RC(c_i)~\lor (closed(c_i) \wedge closed(c_j))$;
                    \State $RC(c_j)=RC(c_j)~\lor (closed(c_i) \wedge closed(c_j))$;
                \EndIf
            \EndWhile
            \State $RC(K)=RC(K)~\wedge~RC(c_i)$;
        \EndIf
    \EndFor
    \EndProcedure
    
    \hspace{1pt}
    \setcounter{ALG@line}{0}
    
    \Procedure{is\_comb\_cycle}{sub\_circuit $S$}
        \State $r,~r' \leftarrow$ input and output of auxiliary-circuit;
        \If{$SAT(S_{opened} \wedge (r \neq r'))$}
            \State return False;
        \Else
            \State return True;
        \EndIf
    \EndProcedure
\end{algorithmic}
\end{algorithm}

For this purpose, during the pre-processing step, in addition to composing the "no sensitizable path" clauses (as proposed in \cite{8203759}), we compose and include a new set of clauses that consider "reducibility" as an alternative option to opening the loops. In this picture, the cycle could either be opened (using no sensitizable path clauses) or could be reduced using newly added reducible clauses. The \textit{reducible} clauses are defined for possible dependent cycles that implement specific functions between their inputs and outputs. These clauses will be generated for each cycle by pairing it with matching outer cycles. The process of generating the reducible clauses is captured in the Algorithm \ref{nested_alg}. The reduction attack procedure, first, sorts all cycles according to their length and then begins processing them from the shortest to the longest cycle. For each cycle, it checks if the cycle is combinational if it is not, it tries to find an outer cycle that makes its behavior combinational. In this algorithm, the $IS\_COMB\_CYCLE()$ validates if a sub-circuit containing a cycle is combinational or not. For this purpose, the function disconnects the cycles by breaking the feedback into two disconnected wire segments $r$ and $r'$. Then by using a SAT solver, it checks if there are any values for the wires that $r \neq r'$. If such a scenario was not found, it classifies the sub-circuit as a combinational circuit. Otherwise, a non-combinational circuit, according to which the necessary clauses are generated.

This algorithm could be applied to any netlist obfuscated using the auxiliary-circuit such as the one in Fig. \ref{dependcycle_break}.c. This circuit has two cycles $c_1=$\{X2\} and $c_2=$\{X1, X2\}. The smallest cycle $c_1$ is oscillating and oscillates when X1 output is 1 as shown in Fig. \ref{dependcycle_break}.d. By considering this cycle as closed and pairing it with its only outer cycle $c_2$ we will have $RC(c_1)=k_0'\lor(k_0\wedge k_1)$. The outer cycle $c_2$ as shown in Fig. \ref{dependcycle_break}.e is also non-combinational and the reducible clauses will be $RC(c_2)=k_1'\lor(k_0\wedge k_1)$. Thus, by closing both cycles, as shown in Fig. \ref{dependcycle_break}.b it can be derived that $r' = r \oplus r' \oplus r' = r$ and the circuit does act as a buffer with no oscillation. The reducible clause for this circuit will be $RC(K)=(k_0'\lor(k_0\wedge k_1)) \wedge (k_1'\lor(k_0\wedge k_1))$ for closing both cycles or opening both cycles since non of them has combinational behavior independently.

It should be noted that these auxiliary-circuits could be in the form of partially intercepted cycles, where more than one outer cycle is partially intercepted with another outer cycle. We acknowledge that for partially intercepted cycles, our proposed algorithm would not work, and an alternative algorithm that generates the \textit{NC} condition by considering the partially intercepted combinational cycles is required.

\vspace{-10pt}
\subsection{Breaking Hard Cycles} \label{breaking_hard_cycles}
\begin{figure}
\centering
    \includegraphics[width=0.65\columnwidth]{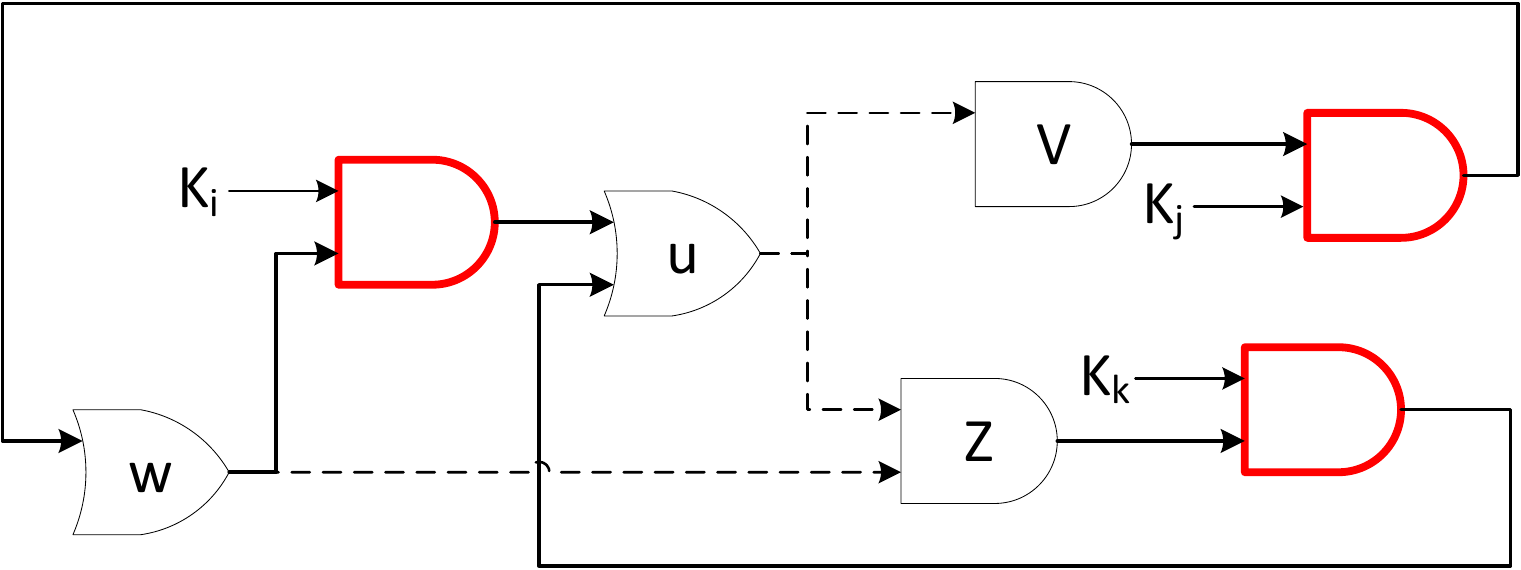}
    \caption{An example for a circuit obfuscated with a hard cycle. Added Key-gates are shown in red, and the riginal wires are shown with dotted lines.}
    \label{hardcycle}
      \vspace{2pt}
\end{figure}

Hard cycles were proposed in \cite{ZhouCycles} to create a situation that any traversal of feedback signals will miss a cycle. An example is shown in Fig. \ref{hardcycle}, where the original circuit consists of gates U, V, W, and Z. In the obfuscated netlist, the gate U is connected to V and Z, and W is connected to Z. By creating a hard cycle, new connections using AND gates have been added. These new wires connect (V, W), (W, U) and (Z, U) and shown with thicker lines. Feedback sets for the new circuit are \{V, W\} and \{Z, U\}. Application of CycSAT attack on this circuit misses the larger cycle \{U, V, W, Z, U\}, and the attack fails.

% By modifying CycSAT attack and by extracting the list of cycles in the pre-processing step, \{U, V, W, Z, U\} will be found and the necessary clauses will be added. So, by adding a simple cycle that would be missed during the original attack, cyclic obfuscation would not be feasible and the circuit could be still deobfuscated.

% \subsection{CycSAT Runtime Dependents on the Number of Cycles}\label{breakingCycSAt}

% By looking at the netlist as a flow-graph, CycSAT \cite{8203759} first finds a list of feedback signals $W = (w_0, w_1, ...w_m)$ whose break make the circuit acyclic. The list of such feedback signals could be found by sorting the nodes in a topological order using a BFS traversal of the graph. During the BFS search, whenever the next node of BFS is a node that is previously visited, its connecting edge is marked as a feedback. Based on graph theory, this could be done in O(V+E) where V is the number of gates, and E is total number of fanouts in a design.

% In the next step, Cyclic SAT assures that there is no structural path from a feedback to itself. This is done by starting from a feedback edge, and constructing the "no structural path condition" in the netlist until the feedback is reached again when completing the loop. The no structural path condition is tested using function $F(w_i,j)$, which is defined as: 

% \begin{equation} \label{eq:erl}
% F(w_i,j) = \bigwedge_{l\in NK(j)}F(w_i,l)\lor bk(l,j)
% \end{equation}

% In this case, $NK(j)$ is the set of non-key fanins of signal $j$, and $bk(l,j)$ is the condition on the key ensuring that $l$ (a fanin wire) does not affect $j$. The $F(w_i,j)$ starts with the condition:

% \begin{equation} \label{eq:erl}
% F(w_i,j) = 0, \forall i \in W 
% \end{equation}

Hard cycles could be easily broken by modifying the mechanisms used for the computation of $F(w_i,j)$. The $F(w_i,j)$ could be computed in two ways: (i) traversing through a cycle starting from $w_i$ until $w_i$ is visited again and ignoring the cycle break conditions imposed by fanins of other nested cycles; or (ii) traversing through one cycle and adding the cycle break conditions imposed by other nested cycle. As shown for the example in Fig. \ref{hardcycle} and the next example, the first choice results in missing some "no cycle" (NC) conditions, leaving cycles in a design that could break the subsequent SAT attack. By choosing the condition (ii), we show that it is possible to build the \textit{NC} condition by visiting all cycles in the netlist without missing any of the hard cycles. To better illustrate this concept, consider the following example: For the obfuscated netlist in Fig. \ref{graphloop} and a topological sort from gate A, the edge $E$ and $F$ are identified as feedbacks. When following rule (i), and after building the \textit{NC} condition we will have:

\begin{figure}[t]
    \centering
    \subfloat[]{\includegraphics[width=0.25\columnwidth]{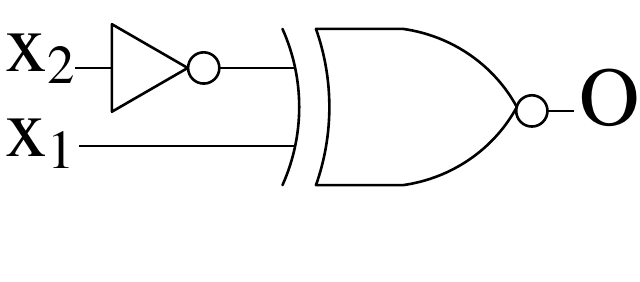}}
    \hspace{1cm}
    \subfloat[]{\includegraphics[width=0.3\columnwidth]{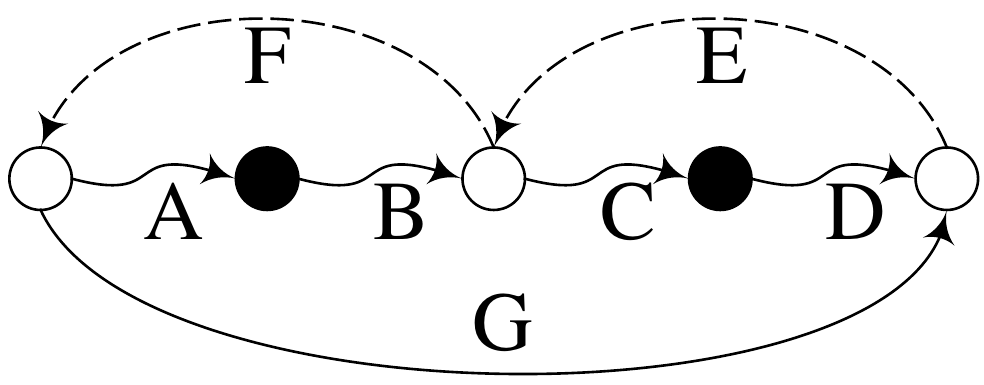}} \\
    \subfloat[]{\includegraphics[width=0.6\columnwidth]{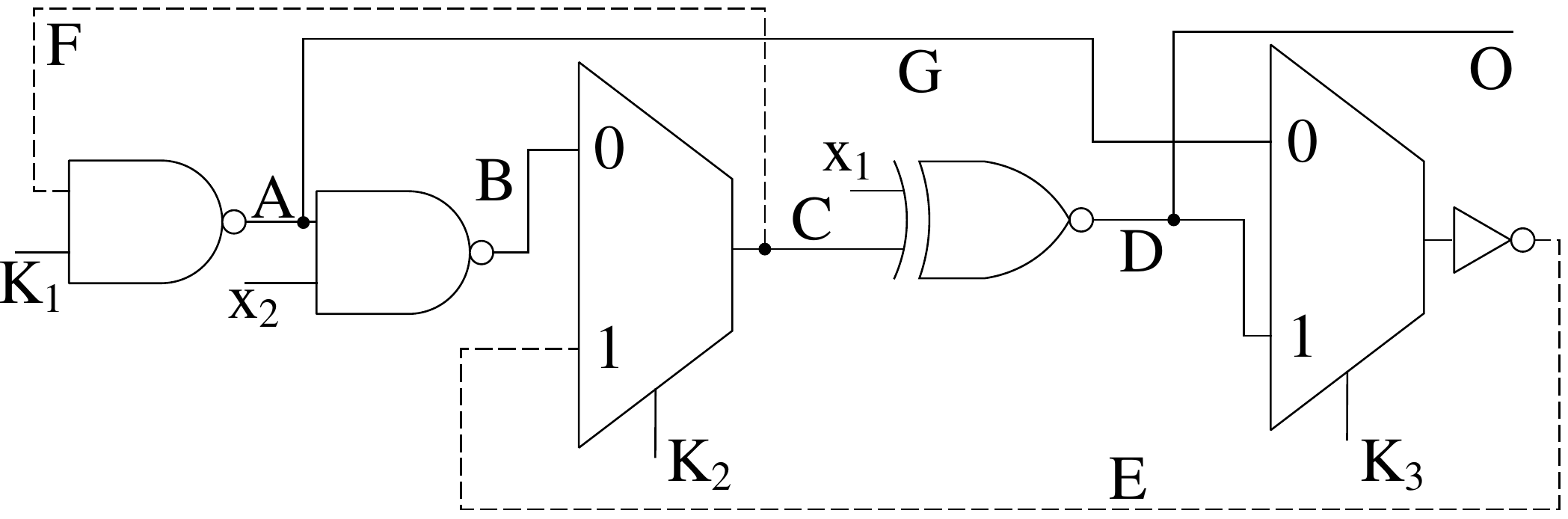}}
    \caption{(a) The Original circuit (b) The flow diagram of the obfuscated circuit (c) The Cyclically Obfuscated circuit. \vspace{3mm}}
    \label{graphloop}
\end{figure}

\begin{figure}[H]
\begin{algorithmic}[1]
\scriptsize
\State $F(F,A) = F(F,F) \lor bk(k_1) = k_1'$ 
\State $F(F,F') = F(F,A) \lor bk(k_2) = k_1' \lor k_2$
\State $F(E,C) = F(E,E) \lor bk(k_2) = k_2'$
\State $F(E,E') = F(E,C) \lor bk(k_3) = k_2' \lor k_3'$
\State $NC=F(F,F')\wedge F(E,E') = (k_1' \lor k_2) \wedge (k_2' \lor k_3')$
\end{algorithmic}
\vspace{-5pt}
\end{figure}

The problem with this assignment is when $(k_1,k_2,k_3)=(1,1,0)$. In this case, the \textit{NC} condition is satisfied, however, the larger nested cycle $\{E, F, G, E\}$ is not broken. Hence, the \textit{NC} condition would not resolve the cycles if nested or multi-path scenarios exist. In this case, if the wrong key $(k_1,k_2,k_3)=(1,1,0)$ is chosen by SAT solver, it will enter a loop. Depending on whether the cycle is oscillating or stateful, the SAT solver will either be trapped in an infinite loop or will exit UNSAT. Note that this infinite loop happens during the execution of the SAT solver and not during the topological sort used in the original SAT attack proposed in \cite{7140252,el2015integrated}.

To avoid the problem imposed by rule (i), we need to follow the rule (ii) where the key contribution of all fanins in all stages are considered. When using rule (ii) for building the \textit{NC} condition for the same circuit we have:

\begin{figure}[H]
\begin{algorithmic}[1]
\scriptsize
\State $F(F,A) = F(F,F) \lor bk(k_1) = k_1'$ 
\State $F(F,F') = (F(F,A) \lor bk(k_2)) \wedge (F(F,E) \lor bk(k_2)) = (k_1' \lor k_2) \wedge (k_1' \lor k_3 \lor k_2'$) 
\State $F(E,C) = F(E,E) \lor bk(k_2) = k_2'$
\State $F(E,E') = (F(E,C) \lor bk(k_3)) \wedge  (F(E,G) \lor bk(k_3))  = (k_2' \lor k_3') \wedge (k_2' \lor k_1' \lor k_3)$
\State $NC=F(C,C')\wedge F(E,E') = (k_2' \lor k_3') \wedge (k_1' \lor k_2' \lor k_3) \wedge (k_1' \lor k_2).$
\end{algorithmic}
\vspace{-5pt}
\end{figure}

By following the rule (ii), the previous assignment of keys $(k_1,k_2,k_3)=(1,1,0)$ will no longer be a valid assignment, preventing the SAT solver from being stuck or exiting with a wrong key. However, in this case, \emph{all cycles in the design have to be traversed and conditioned}. As a matter of fact, given the way the \textit{NC} is formulated in \cite{8203759}, to derive the "no structural path" condition, some of the combinational cycles (such as $\{E, F, G, E\}$ in Fig. \ref{graphloop}) have been visited more than once. Hence, the number of times the key conditions have to be generated is even larger than the number of netlist cycles.

%\textbf{Lemma.} \emph{When building "no structural path" based on rule (ii), all cycles in a netlist the Directed Cyclic Graph are visited.}

%\textbf{Proof}: letگs assume that there is a cycle $C_{missing}$ that is not visited. Given that it is a cycle, it does contain at least one feedback $w_f$ which is used in NC condition generation but with a different choice of cycle, cycle $C_f$ for traversal. Considering that rule (ii) is followed for the NC construction, we conclude that that is no key gate shared between $C_{missing}$ and $C_f$  that share one or more keyed signals with cycle containing this feedback, otherwise no structural path condition on that key-gate would have discovered and traversed this cycle, including signal $w_f.  

The problem of visiting nested cycles more than once in CycSAT attack could be resolved by a slight modification to CycSAT pre-processing step. In the modified attack, instead of applying rule (ii) on one-cycle-per feedback, we could apply the rule (i) on all cycles. It is intuitive to see that both approaches produce the same \textit{NC} clauses. For example, in Fig. \ref{graphloop} when following condition (i), and traversing cycle $\{E, F, G, E\}$, the condition $(k_1' \lor k_2' \lor k_3)$ is generated. Hence, by ANDing the generated condition to the two clauses generated by applying the rule (i), the \textit{NC} condition of rule (ii) is generated. However, in this case, the combinational cycle $\{E, F, G, E\}$ is only visited once. Even by considering the improvement suggested in CycSAT formulation, it still requires visiting all cycles in a netlist to compose the \textit{NC} clauses. This necessity, as described in the next section, becomes one of the key features which is used in this paper to break CycSAT attack.

A different method of introducing complexity is by eliminating the DAG nature of the original netlist and by transforming it into a Boolean cyclic function, which could be represented using a Directed Cyclic Graph (DCG), before subjecting it to cyclic obfuscation. If the original netlist is not a DAG, CycSAT pre-processing step has to build the \textit{NC} condition by checking for "\emph{no sensitizable path}" condition \cite{8203759}, instead of "\emph{no structural path}" condition. The no sensitizable path condition from \cite{8203759} is recited in equation~\ref{eq:NSenP}: 

\vspace{-7pt}
\begin{equation} \label{eq:NSenP}
\footnotesize
F(w_i,j) = \bigwedge_{l\in fanin(j)}F(w_i,l)\lor ns(l,j)
\end{equation}
\vspace{-10pt}

The "no sensitizable path" condition generates a  clause for each multi-input gate in a cycle. As a result, \textit{NC} clauses are much longer and much weaker. Hence, adding even a small number of feedbacks to such circuits (that have valid Boolean cycles) for obfuscation, will significantly increase the size of the circuit-SAT problem, as the "no sensitizable path" condition has to be generated for all cycles. To illustrate the weaker and longer nature of the \textit{NC} clauses, the "no sensitizable path condition" for the circuit in Fig. \ref{graphloop} is constructed below:

\begin{figure}[h]
\begin{algorithmic}[1]
\scriptsize
\State $F(F,A) = F(F,F) \lor ns(F,A) =\enspace k_1'$ 
\State $F(F,B) = F(F,A) \lor ns(A,B) = k_1' \lor x_2'$
\State $F(F,F') = (F(F,B) \lor ns(B,F')) \wedge (F(F,E) \lor ns(E,F')) = (k_1' \lor x_2' \lor k_2) \wedge (k_1' \lor k_3 \lor k_2'$) 
\State $F(E,C) = F(E,E) \lor ns(E,C) = k_2'$
\State $F(E,D) = F(E,C) \lor ns(C,D) = k_2'$
\State $F(E,E') = (F(E,D) \lor ns(D,E')) \wedge (F(E,G) \lor ns(G,E')) = (k_2' \lor k_3') \wedge (k_2' \lor k_1' \lor x_2' \lor k_3)$
\State $NC=F(F,F')\wedge F(E,E') = (k_1' \lor x_2' \lor k_2) \wedge (k_1' \lor k_3 \lor k_2') \wedge (k_2' \lor k_3') \wedge (k_2' \lor k_1' \lor x_2' \lor k_3)$
\end{algorithmic}
\vspace{-12pt}
\end{figure}

% In the section \ref{cyclock} we propose two methods for building an exponential relation between the number of generated cycles in a netlist with respect to the number of inserted feedbacks and subsequently introduce three techniques for transforming a netlist to contain valid cyclic Boolean functions to  force an attacker to use the "no sensitizable path" condition in the CycSAT attack. 

\section{SRC-Lock: The Proposed Cyclic Obfuscation}\label{cyclock}
The issue with the original method of generating cycle avoidance (NC) clauses using CycSAT was shown and discussed in section \ref{cycobfuscation} using two simple examples in which traversal of wires based on a single topological sort of gates resulted in a missing cycle. When using the original CycSAT, because of the missing \textit{NC} clauses for such cycles and due to the randomness of assigned key and input values by the SAT solver, the SAT attack can be stuck in an infinite loop or exit with a wrong key. The possibility of facing an oscillating or stateful cycle greatly increases as the number of generated cycles in the design increases to a point that majority of key-space (to be tested by SAT solver) could result in oscillating or stateful cycles, vanishing the chances of a successful attack to unreasonably small probability. On the other hand, attacks such as BeSAT \cite{besat} that can track the behavior of the SAT attack at runtime, could detect oscillating or stateful scenarios (due to missing cycles in pre-processing time) and eliminate the incorrect key. However, at runtime, the BeSAT eliminates one key at a time. Hence, it is successful if the number of such key combinations are small. In other words, the BeSAT attack run time is linearly dependent on the number of such keys. When such key combinations are (exponentially) large (which is the case in our to-be-proposed obfuscation solution), the BeSAT attack's runtime becomes unacceptably large.
% This is when, such cycles, with a single cycle avoidance clause (generated at the pre-processing step), could be constrained such that the SAT attack extracts the correct key in a single iteration. However, again, this requires the pre-processing to be done for all cycles.}

CycSAT pre-processing time is characterized in equation \ref{eq:preproceTime}. As illustrated, the processing time is linearly related to the number of discovered cycles $N$ and the time for composing the \textit{NC} condition $t_{NC}$ per cycle. Our approach for breaking CycSAT is to exponentially increase the time needed for composing the \textit{NC} condition in the pre-processing step of CycSAT beyond acceptable. This is achieved by exponentially increasing the number of cycles $N$ in a design with respect to the number of inserted feedbacks $m$, and increasing the time required for processing each cycle ($t_{NC}$) by forcing the pre-processing step to consider the "no sensitizable path" condition instead of "no structural path". Next, we provide two solutions for building an exponential relation between the number of feedbacks and number of generated cycles, and three solutions for converting an acyclic circuit to a valid cyclic circuit.

\vspace{-4pt}
\begin{equation} \label{eq:preproceTime}
T_{NC} = \sum_{i=1}^N t_{NC} \enspace | \enspace  N = 2^{m}
\vspace{-8pt}
\end{equation}

% \vspace{-20pt}
\subsection{Exponentially increasing the number of cycles in a netlist}
In order to exponentially increase the number of cycles in a given netlist with respect to the number of inserted feedbacks, we introduce two approaches: (1) building Super Cycles (SC) and, (2) building Logarithmic Feedback Networks (LFN). \\[-7pt]

\subsubsection{\textbf{Building Super Cycles (SC)}} \label{supercyclesection}
\begin{figure}
    \centering 
    \subfloat[]{\includegraphics[width=0.8\columnwidth]{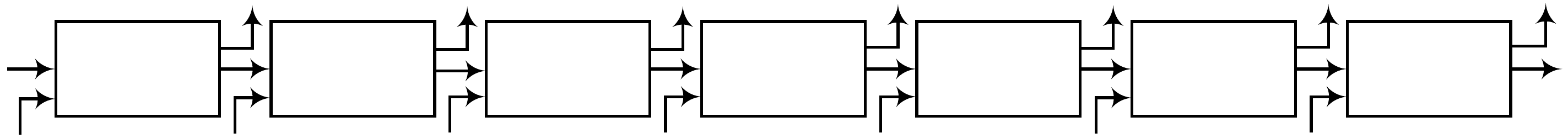}} \\
    \vspace{-5pt}
    \subfloat[]{\includegraphics[width=0.8\columnwidth]{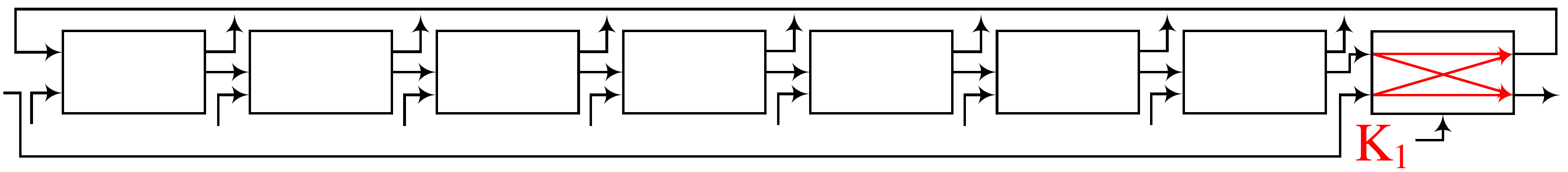}} \\
    \vspace{-5pt}
    \subfloat[]{\includegraphics[width=0.8\columnwidth]{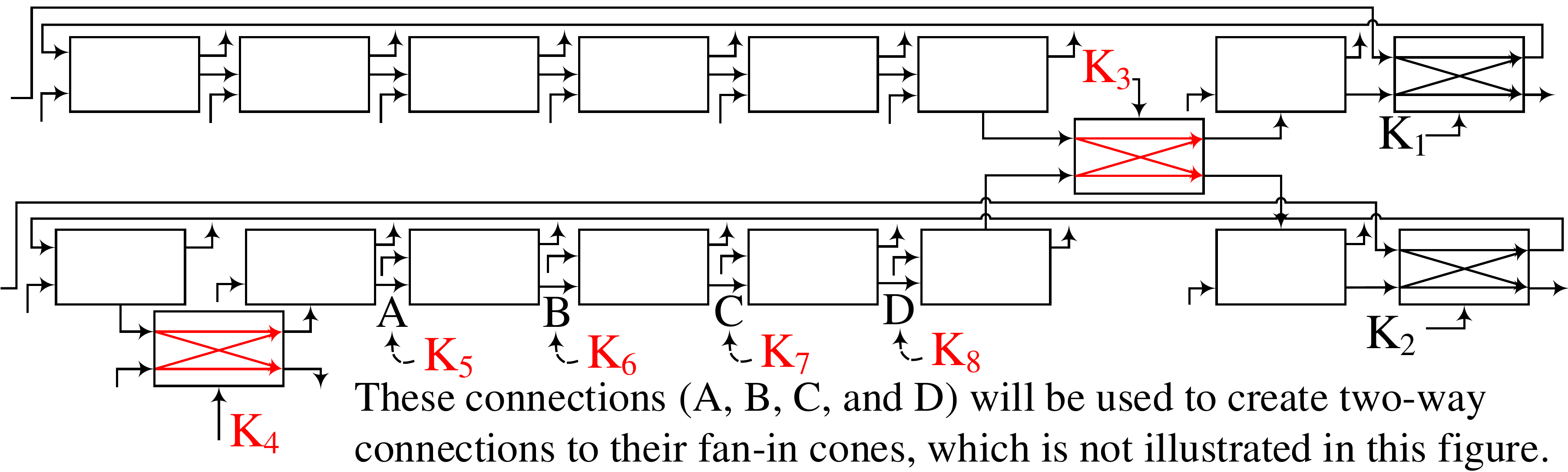}}
    \caption{Building a Super Cycle from 7 gate MC. (a) A path segment containing 7 gates, (b) building a Micro Cycle, (c) building a SC by strongly connecting multiple MCs.}
    \label{buildingSCfig}
\end{figure}

The process of building a SC is illustrated in Fig. \ref{buildingSCfig}. A Micro Cycle (MC) is a cycle created by following the cycle creation conditions adopted from \cite{Shamsi:2017:COC:3060403.3060458}, which are recited below: \\[-7pt]

\noindent
\textbf{MC Condition 1:} Any created cycle has to be non-reducible, \\
\textbf{MC Condition 2:} At least $n\geq2$ edges in each small cycle have to be removable.\\[-7pt]

A reducible cycle has a single entry point. Hence, the depth-first-search (DFS) traversal of a netlist that only contains reducible cycles is unique. This allows the reducible cycles to be easily opened by removing a unique set of feedback edges which can be found efficiently \cite{Shamsi:2017:COC:3060403.3060458}. By having multiple entries into each MC, the non-reducible condition is satisfied, forcing an adversary to use CycSAT pre-processing step to generate the necessary cycle avoidance clauses before invoking the SAT solver. In graph theory, a strongly connected graph is defined as a graph with at least one path between any two pairs of its vertices. adopting from this definition, in our solution, a SC is defined as a strongly connected graph of MCs. To substantially increase the number of generated cycles, in the last step of SC generation, the edge density of the generated strongly connected graph is increased, creating additional paths between MCs. The process of building a SC is summarized in Algorithm \ref{buildingSC}. 

% \begin{figure}
% \centering
%     \includegraphics[width=120pt]{images/switch.pdf}
%     \caption{Switch structure.}
%     \label{switch}
%     %  \vspace{-8pt}
% \end{figure}

% Both approaches use a switch to create a direct or cross connection between two points which is shown in Fig. \ref{switch}. Switch inputs will be defined by SC and LFN methods and a single key input will determine connections between input and output.

\begin{algorithm}
\caption{Steps for building a Super Cycle \label{buildingSC}}
\begin{algorithmic}[1]
\footnotesize
\State Construct MCs in the fanin of smallest possible number of primary outputs.
\State Strongly connect all generated MCs (this, as illustrated in Fig. \ref{buildingSCfig}.b, is done by creating a two-way connection between each newly created MC, and the existing SC).
\State Select signals in MCs (A, B, C, D in Fig. \ref{buildingSCfig}.c) that are not used for SC connectivity and provide a two way path from them to unused edges in other MCs or random signals in their fanin cone.
\end{algorithmic}
\end{algorithm}

In this algorithm, the requirement of generating the MCs in the fanin of the smallest number of primary outputs increases the likelihood of shared and/or connecting edges between created MCs. By having all MCs strongly connected, we create the possibility of larger combinational cycles. And finally, adding the random connections, increase the density of the edges in the strongly connected graph, increasing the number of resulting cycles. In the results section, we illustrate that the number of created cycles, generated from following these steps as described in Algorithm \ref{buildingSC}, becomes an exponential function of the number of inserted feedbacks.\\[-7pt]

\textbf{Lemma.} \emph{The lower bound on the number of cycles created when using SC is $2^m$, when $m$ is the number of inserted feedbacks.}

\emph{Informal Proof.} The proposed SC method adds two paths (from and to paths) to connect each new cycle to the existing SC. This way, the new cycle could be added or not added to any of the previously existing cycles. Hence, the addition of a new cycle at least double the number of potential cycles. Note that the number of connecting edges between the new cycle and the existing cycle could be more than one, resulting in an increase in the number of cycles with a much higher rate. From this discussion, after inserting $m$ feedbacks and connecting them, at least $2^m$ cycles will be created. $\blacksquare$ \\[-7pt]

\subsubsection{\textbf{Building Logarithmic Feedback Networks (LFN)}}\label{buildingLFN}
\begin{figure}
    \centering
    \subfloat[]{\includegraphics[width=0.7\columnwidth]{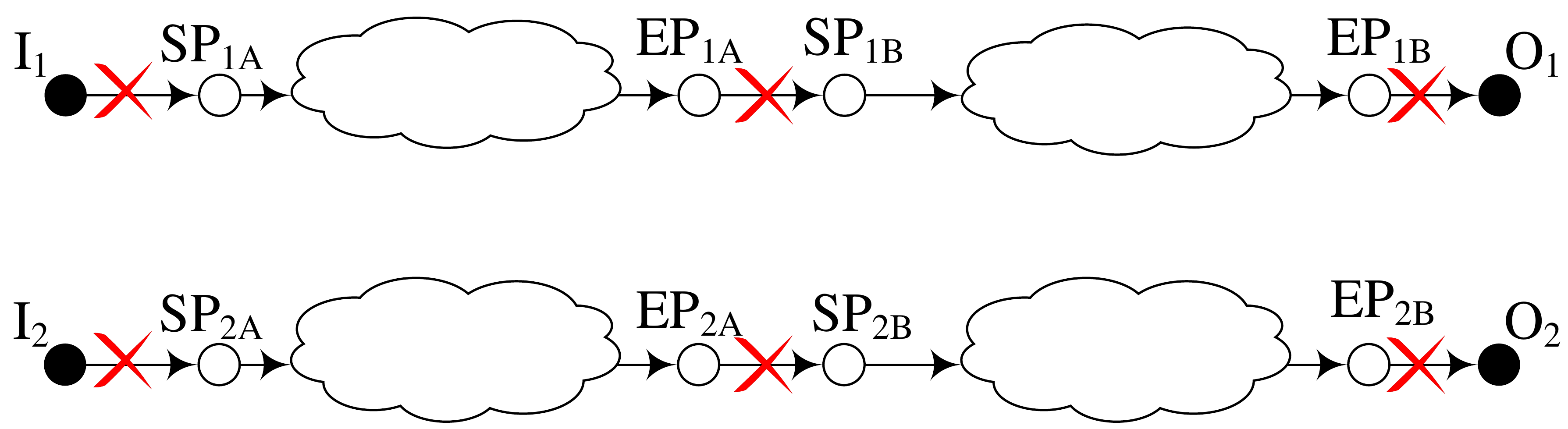}} \\
    \subfloat[]{\includegraphics[width=0.40\columnwidth]{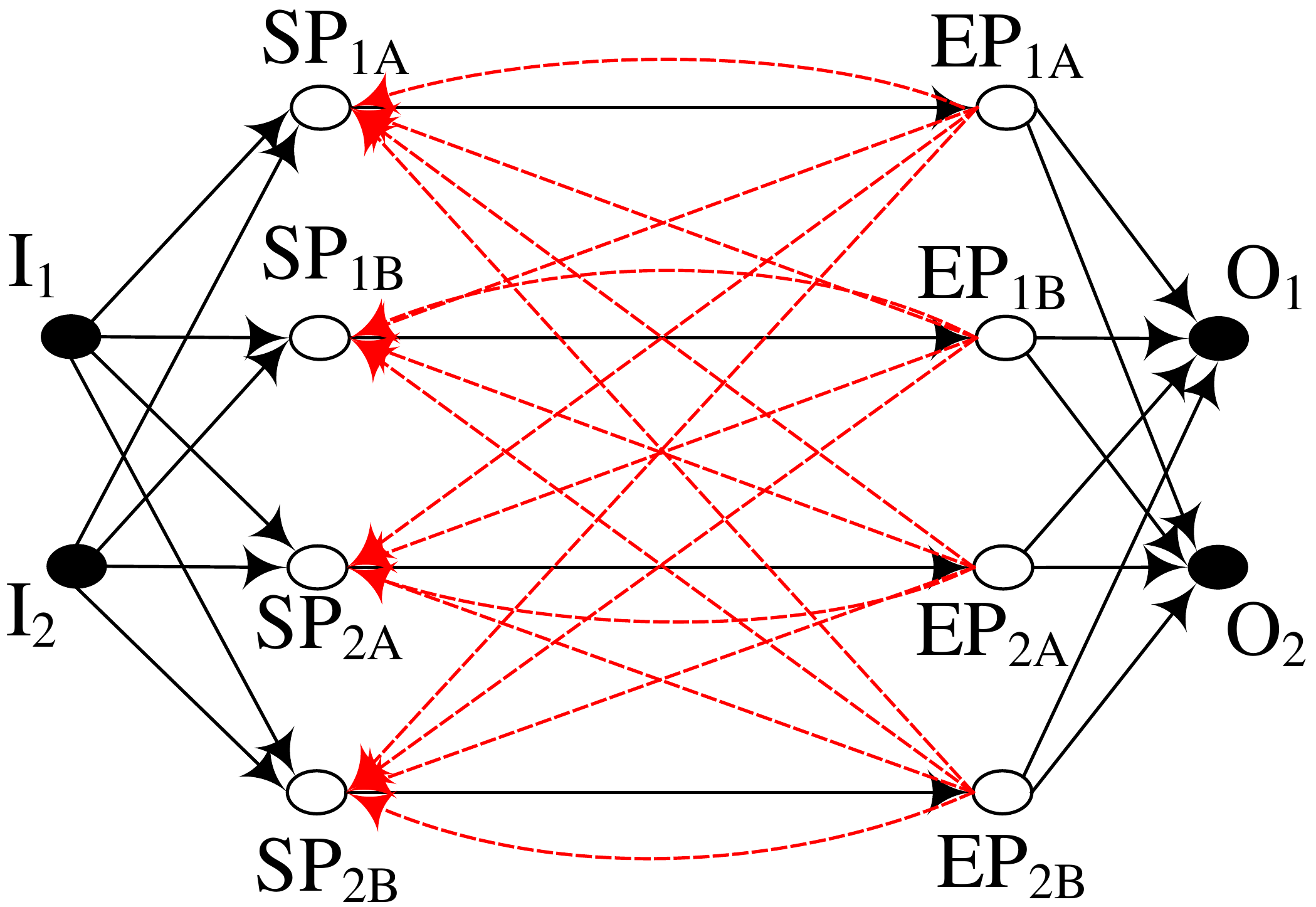}}
    \hspace{0.5cm}
    \subfloat[]{\includegraphics[width=0.42\columnwidth]{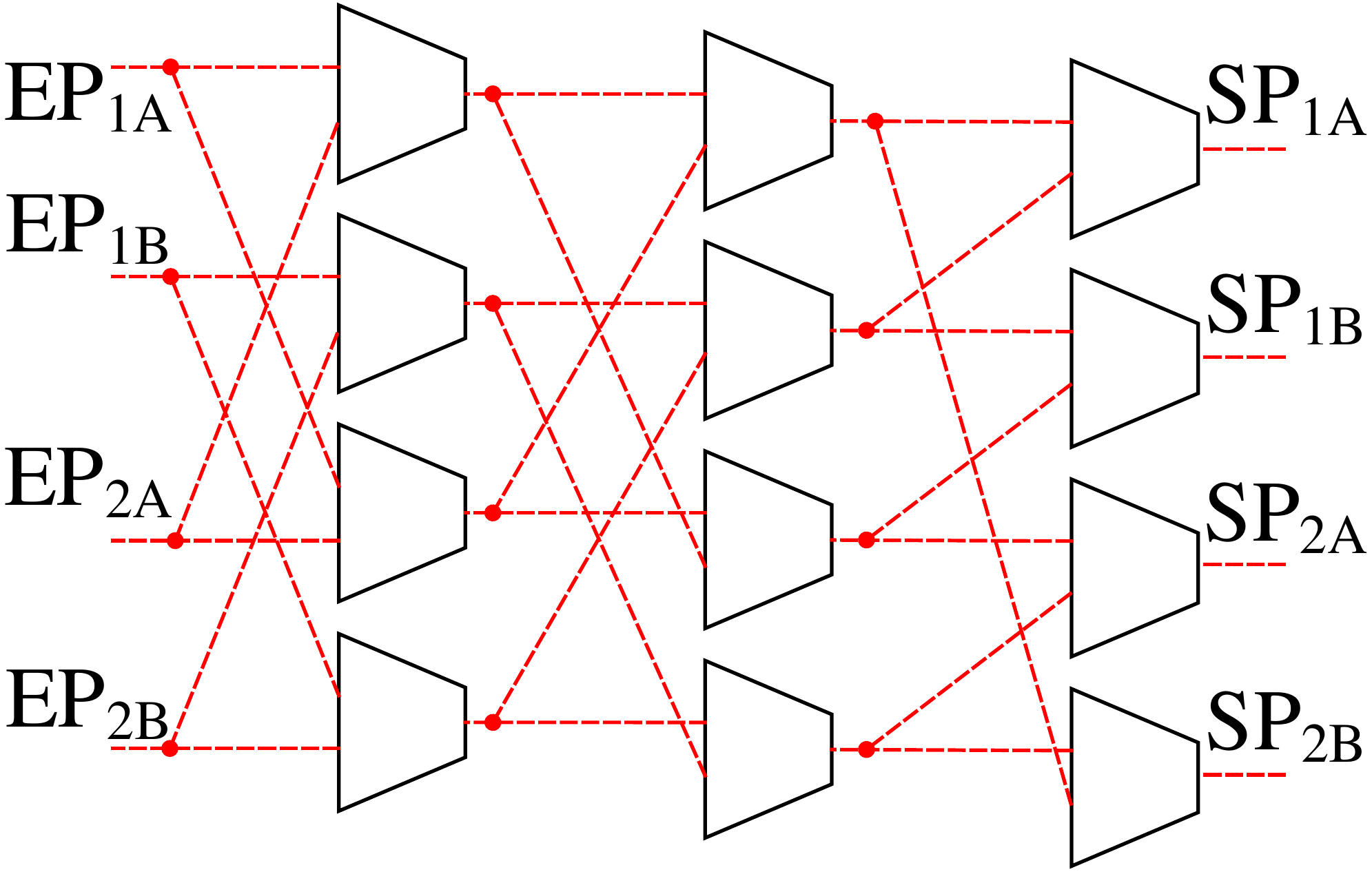}}
    \caption{Building a logarithmic feedback network in which the number of cycles exponentially increase with the number of feedbacks. \vspace{4mm}}
    \label{logNet}
\end{figure}

In this method, as illustrated in Fig. \ref{logNet}.a, several logic paths (preferably from the fanin cone of a single primary output) are selected. Then, by breaking a wire in the midpoint of each logic path, we create two smaller logic segments. The signal entering and the signal exiting each half segment is marked as its start point (SP) and endpoint (EP) respectively. Then, the SP and EP of multiple such logic path segments are used to build a logarithmic switching network (e.g., Omega, Butterfly, Benes, or Banyan network). When connecting $M$ number of EPs to $M$ number of SPs, for $M$s of power of 2, we need  $M(1+log_2(M))$ multiplexers for a logarithmic network. In this case, when the correct key is applied, the switching network is configured correctly, otherwise, invalid connectivity obfuscates the netlist functionality.

\textbf{Lemma.} \emph{The lower bound on the number of cycles created when using LFN is $\sum_{l=1}^m {m \choose l}(l - 1)!$, when $m$ is the number of inserted feedbacks and $l$ is the cycle size divided by 2.}
% log base two of the number of cycles of size $l$.}

\emph{Informal Proof.} The proposed LFN is a special case of a complete bipartite graph that contains no odd cycles. Suppose that $SE_{ij}$ indicates a vertex from $SP_i$ to $EP_j$. Similarly, $ES_{ij}$ indicates a vertex from $EP_i$ to $SP_j$. For $l = 2$, the cycles are all paths from a $SP$ to its corresponding $EP$ and return path $\{SE_{ii}, ES_{ii}\}$. If we start from $SP_{i}$, the second visited node is its $EP$ ($EP_i$). Since each $EP$ is connected to all $SP_s$, for intermediate nodes, we have all permutations as alternative possible paths. Cycles with $l = 2$, have no intermediate node. So, there are ${m \choose 1}0!$ cycles when $l = 2$. For $l = 4$, the cycles are paths like $\{SE_{ii}, ES_{ij}, SE_{jj}, ES_{ji}\}$. There is only one intermediate node in cycles when $l = 4$ resulting in ${m \choose 2}1!$ cycles. Similarly, for $l = 6$, the cycles are paths like $\{SE_{ii}, ES_{ij}, SE_{jj}, ES_{jk}, SE_{kk}, ES_{ki}\}$. Since, we have two intermediate nodes, $j$ and $k$, we should consider their permutation as a new cycle, i.e. $\{SE_{ii}, ES_{ik}, SE_{kk}, ES_{kj}, SE_{jj}, ES_{ji}\}$. So, for $l = 6$, we have ${m \choose 3}2!$. With similar relation, for $l = 8$, we have ${m \choose 4}3!$ cycles. We can extend this relation to all cycles with different length. The summation of these cycles indicates the number of cycles in our logarithmic network, which is $\sum_{l=1}^m {m \choose l}(l - 1)!$.~$\blacksquare$ \\[-7pt]

Note that $\sum_{l=1}^m {m \choose l}(l - 1)!$ is the lower bound of the number of simple and nested cycles created by using the logarithmic network. The number of paths from each SP to each EP could be more than 1, and there are possibilities of having a connection between SPs and EPs of the different paths in the original circuit, increasing the number of cycle possibilities to a far larger number. Based on the lower bound formula, the number of created cycles is $O(\sum_{l=1}^m {m \choose l}(l - 1)!) \leq  O(m!) = O(m^m)$. Hence, there exists an exponential relation between the number of inserted feedbacks and the number of resulting cycles in the netlist.

 \vspace{-5pt}
\subsection{Building Cyclic Boolean Functions}
A Boolean function does not need to be acyclic. Furthermore, it is possible to reduce the number of gates in a circuit if a function could be implemented in its acyclic form \cite{7406959,1466160,1218927,Rivest:1977:NFM:1310165.1310794}. For example, the work in \cite{Rivest:1977:NFM:1310165.1310794} presents an n-input 2n-output positive unate Boolean function which can be realized with $2n$ two-input gates when feedback is used but requires $3n-2$ gates if the feedback is not used. Hence, cyclification of a circuit in addition to forcing CycSAT pre-processing step to consider the "no sensitizable path", could also remedy the area overhead of introducing new gates for cyclic obfuscation. To cyclify a netlist and to increase the $t_{NC}$ in Equation \ref{eq:preproceTime}, we suggest three approaches: (1) Template-based cyclic-function mapping, (2) Input-dependency based cycle generation and, (3) Node-merging cycle generation. \\[-7pt]

\begin{figure}
\centering
    \includegraphics[width=0.6\columnwidth]{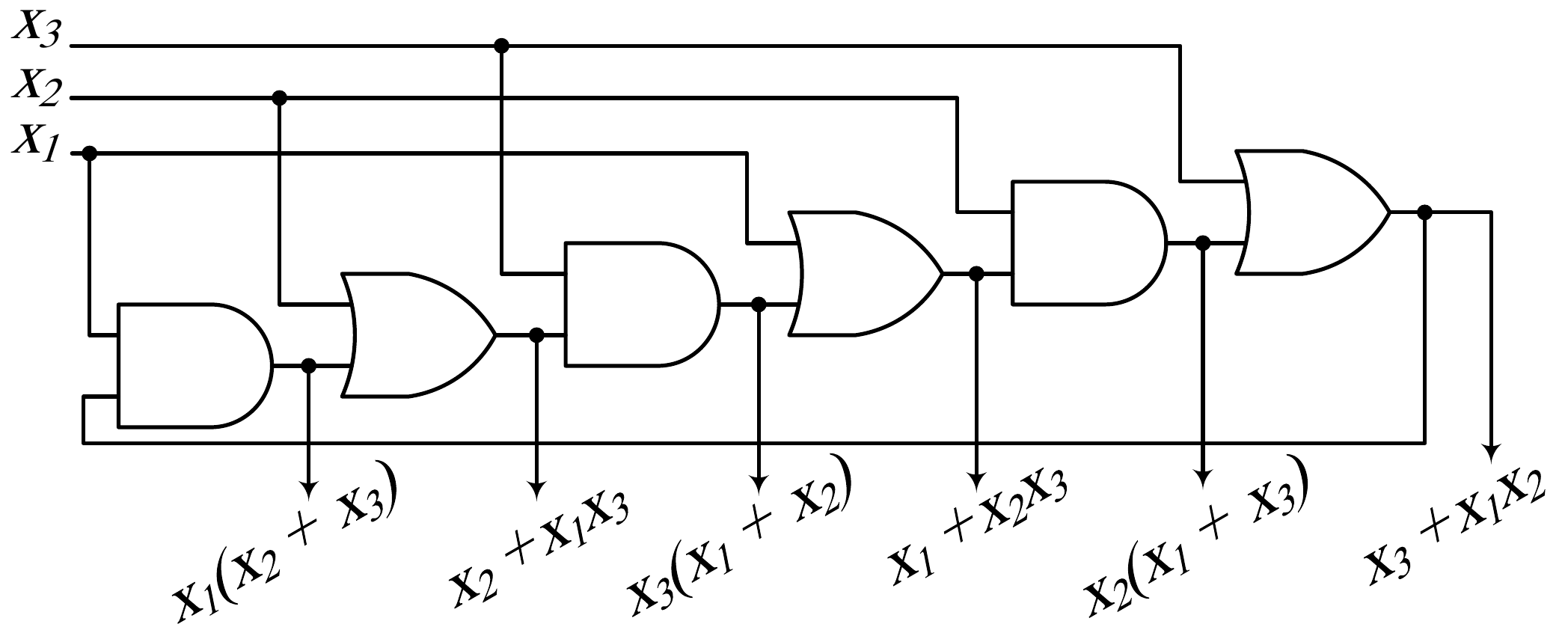}
    \caption{3-input Rivest circuit implementing six functions.}
    \label{Rivest}
    % \vspace{-10pt}
\end{figure}

\subsubsection{\textbf{Template-based cyclic-function mapping}}
In this approach, many small cyclic Boolean circuits are collected as templates in our obfuscation library. Then, a netlist is scanned for opportunities (with and without logic manipulation) to replace a cluster of logic gates with such templates. An example of such feedback template is the circuit introduced in~\cite{Rivest:1977:NFM:1310165.1310794} where a special case of it (for 3 inputs) is illustrated in Fig. \ref{Rivest}. To introduce cycles, the circuit could be modified to introduce at least one of the possible functions in this circuit. The candidate logic cluster is then replaced by the template. To prevent template scanning and removal attacks, in a subsequent camouflaging step (using the gate and route obfuscation) the template will be hidden. Note that many such templates could be made \cite{7406959,1466160,1218927,Rivest:1977:NFM:1310165.1310794}, and by not knowing the template type and the camouflaged technique used to hide the connection, an attacker can not identify and remove these templates.\\[-7pt]

\begin{figure}
\centering
    \includegraphics[width=0.6\columnwidth]{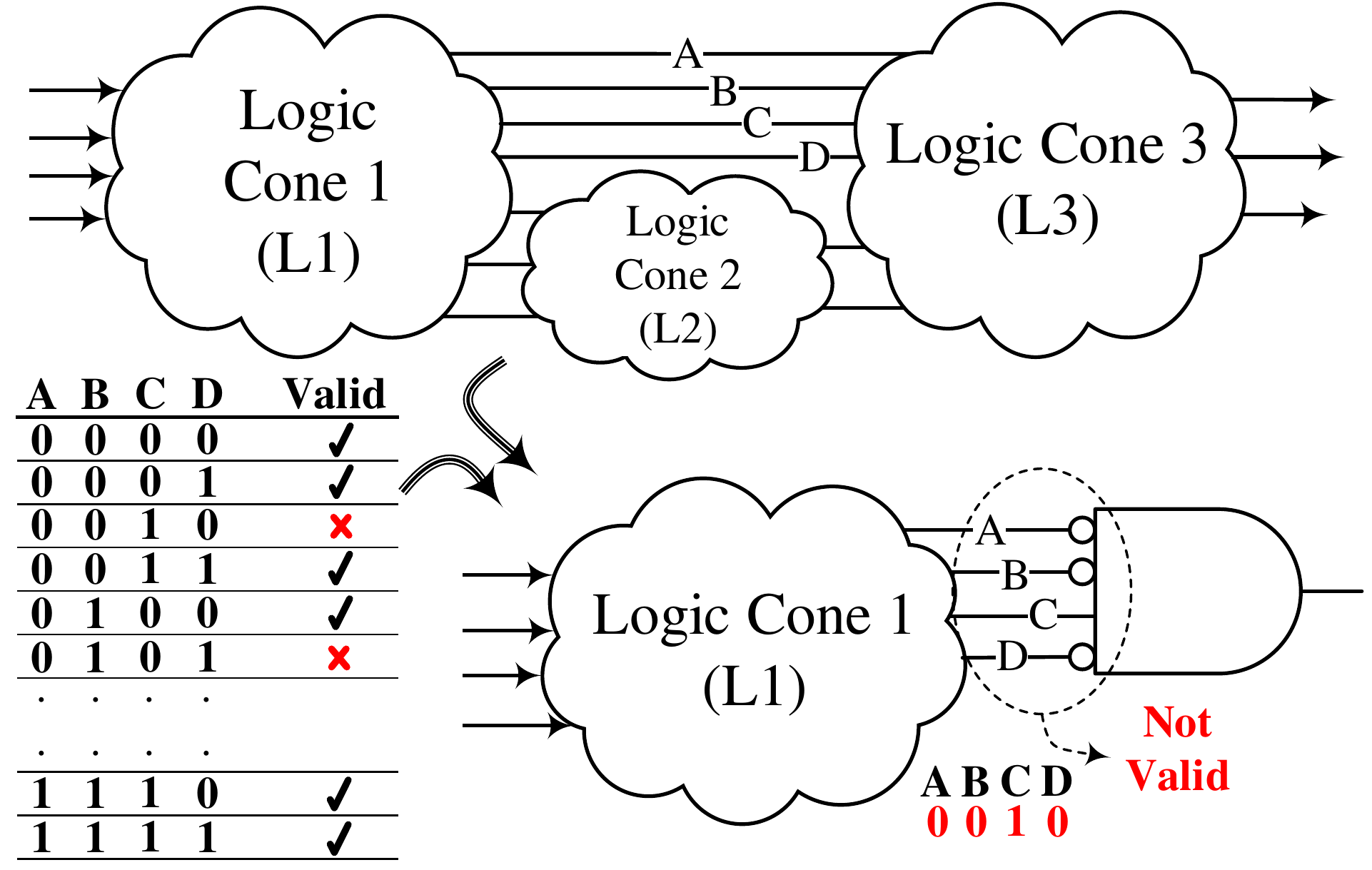}
    \caption{Due to correlation of intermediate signals, certain signal combinations may never occur.}
    \label{SigDepend}
    % \vspace{-10pt}
\end{figure}

\begin{figure}[t]
\centering
    \subfloat[]{\includegraphics[width=0.4\columnwidth]{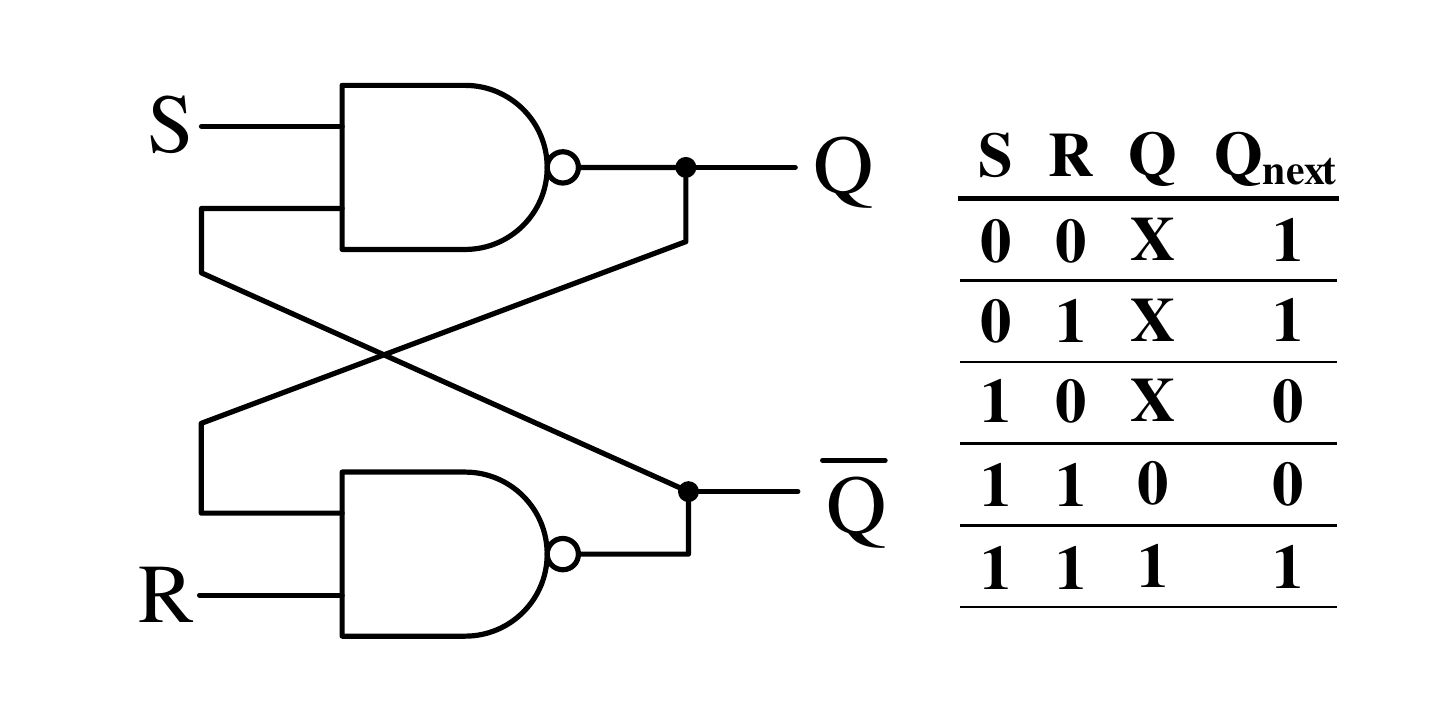}}
    \hspace{.5cm}
    \subfloat[]{\includegraphics[width=0.4\columnwidth]{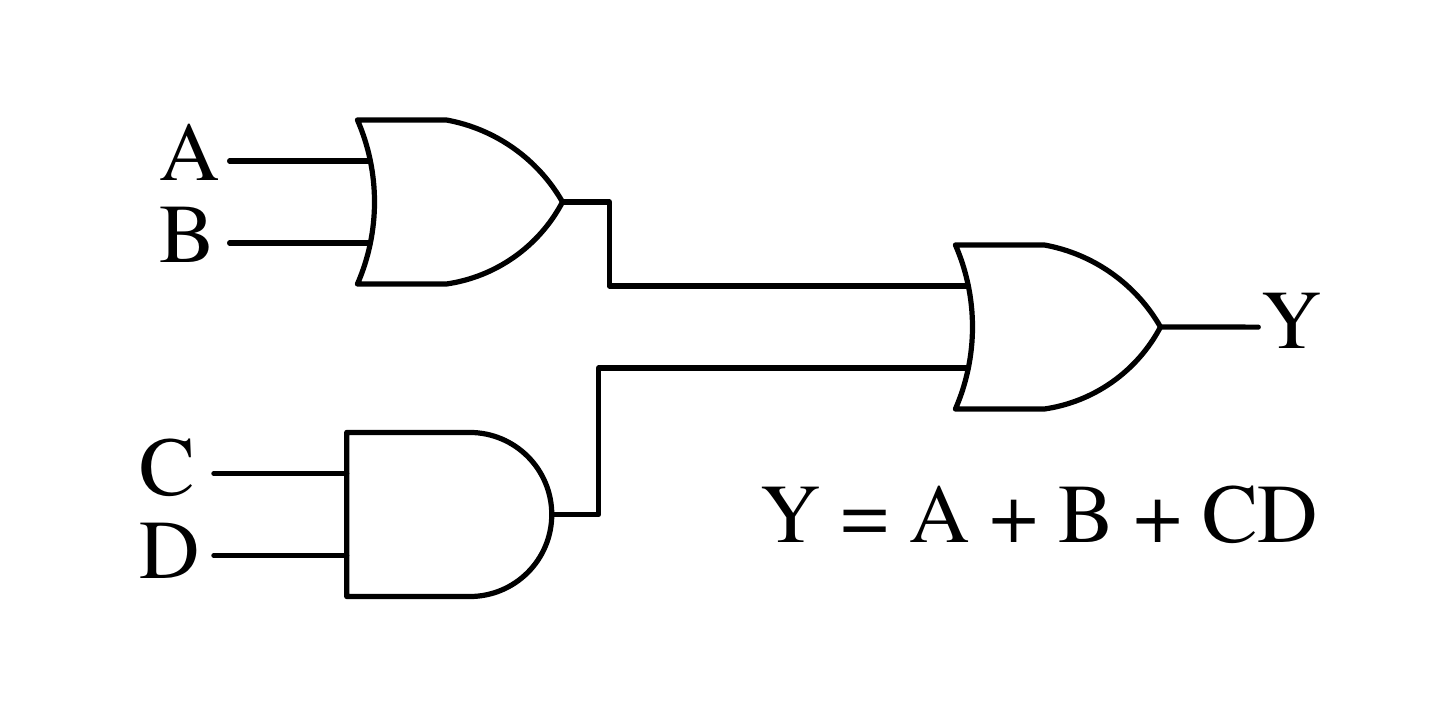}} \\
    \vspace{-5pt}
    \subfloat[]{\includegraphics[width=0.35\columnwidth]{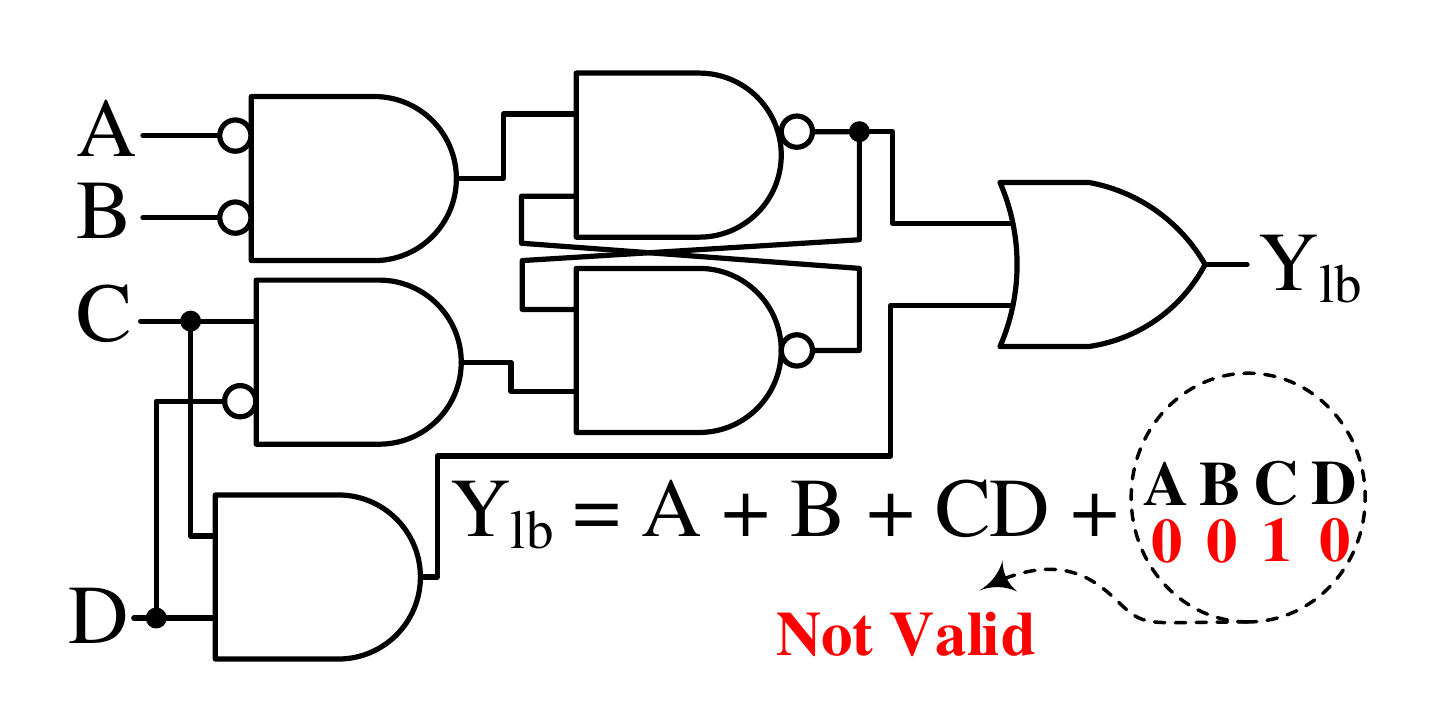}}
    \hspace{.2cm}
    \subfloat[]{\includegraphics[width=0.5\columnwidth]{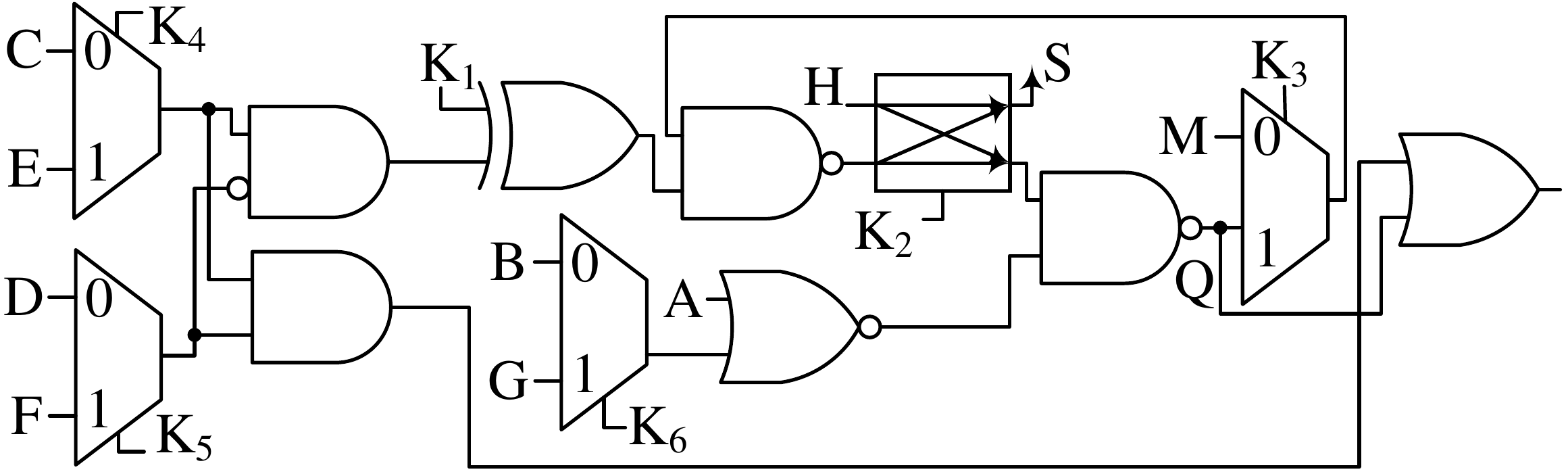}}
    \caption{Input-dependency-based cyclification of a Boolean function. (a) SR-latch (b) original circuit (c) cyclified circuit when $ABCD=0010$ is non-occurring. (d) obfuscated cyclified circuit using additional random inputs $E,F,G,H$ and $M$.}
    \label{latchobf}
\end{figure}

\subsubsection{\textbf{Input-dependency based cycle generation}} \label{idbo}
This method explores the correlations between signals that share common primary inputs in their fanin cone. Considering N such signals in an arbitrary stage of a DAG, some of the $2^N$ inputs may never occur. For example, when tracking 4 signals $A$, $B$, $C$, and $D$ in Fig. \ref{SigDepend}, we may find that $ABCD=\{0010\}$ could not occur. A SAT solver could be used for finding the non-occurring input scenarios; This process is illustrated in Fig. \ref{SigDepend}, where the logic clusters L2 and L3 are removed, and the 4 signals are ANDed together such that for a certain case, for example, $ABCD=0010$, the output of AND gate is evaluated to 1. Then, this circuit is given to a SAT solver to find a satisfying input assignment. If SAT solver returns UNSAT, this combination of input is chosen since it would never happen, otherwise, a different combination is checked.

In the next step, we use a sequential element and tie the discovered non-occurring input scenario to the state preserving input of the sequential element. For example, by using an SR-latch in Fig. \ref{latchobf}.a, If $SR=11$ doesn't happen, the $Q_{next}$ is the inverse of input $S$. Hence, we can build a circuit that ties the discovered non-occurring input scenario to the $SR=11$. For example, let's assume wires A, B, C and D have a non-occurring combination $ABCD=0010$ and these signals construct the signal $Y=A+B+CD$. Fig \ref{latchobf}.c illustrates the signal $Y$ reconstructed when the non-occurring combination of the inputs is tied to SR input of the latch. After generating the cyclic logic, to hide the correlation between input signals, the wire selection is obfuscated. Finally, the SR-latch feedback is obfuscated using a set of multiplexers. This assures that CycSAT can only generate the correct \textit{NC} clauses if the "no sensitizable path" condition is processed, otherwise, it breaks the SR-latch feedback and invalidates the netlist.\\[-7pt]

\subsubsection{\textbf{Node-merging based cycle generation}}
The third approach for cyclification of a netlist is based on the work in~\cite{7406959} where the logic implication is used to identify cyclifiable structure candidates directly, or to create them aggressively in circuits. At its core, the work in \cite{7406959} introduces active combinational feedback cycles by merging two nodes in the original DAG. To check the validity of the generated cyclic netlist, they use a SAT-based algorithm and validate whether the formed cycles are combinational or not.

\section{Timing Aware Cyclic Obfuscation}\label{timawareobf}
During logic locking, each modification to the original netlist affects the timing characteristics of the original circuit. A timing oblivious obfuscation solution could result in changes to the delay of one or more timing critical path(s) (via insertion of key gates), leading to a slower design. In this section, we argue that our proposed obfuscation solution could be designed to be timing-aware, minimizing (or removing) the impact of obfuscation on circuit timing. This can be achieved by incorporating a simple static timing analysis (STA) in our obfuscation procedure. 

% Many key gates, specially if the specified percentage overhead of the targeted cyclic obfuscation is small, could be inserted without affecting the timing. However, after a certain point, any additional key gates would pose a timing penalty for lack of sufficient timing slack in all usable timing paths. In this case, the obfuscation would pose a timing penalty. If the timing penalty is not acceptable, the percentage overhead of cyclic obfuscation should be reduced until the obfuscation is realized. If the timing penalty is acceptable, the additional timing slack will be added to all timing paths, generating new insertion opportunities. 

\begin{algorithm}[t]
\caption{Timing Aware Cyclic Obfuscation}
\label{switchinsertionalgo}
\begin{algorithmic}[1]
\footnotesize
\Procedure{switch\_insertion}{$int~required\_paths$, $circuit~K$}
    \State largest\_cone $\leftarrow$ output port with largest cone;
    \State b = BFS(largest\_cone);
    \While {(number of inserted feedbacks $<$ required\_paths)}
        \State tail $\leftarrow$ pop(b);
        \If{ (slack(tail) $>$ delay of a keygate and tail not marked) }
            \State path $\leftarrow$ DFS on tail considering slacks;
            % \State p $\leftarrow$ selected path;
            \State mark path as selected in the circuit;
            \State add feedback to the path;
            \State update the circuit's timing using STA/EDA;
        \EndIf
    \EndWhile
    
    \For {(each selected path)}
        \For {(each gate in the path)}
            \If{(slack(gate) $>$ delay of a multiplexer)}
                \State disconnect gate output;
                \State insert multiplexer;
                \State connect gate and multiplexer based on SC/LFN;
                \State update circuit timing using STA/EDA;
            \EndIf
        \EndFor
    \EndFor
\EndProcedure

\end{algorithmic}
\end{algorithm}

% \begin{algorithm}
% \caption{Path Selection for Creating Cycles}
% \label{pathselectionalgo}
% \begin{algorithmic}[1]
% \footnotesize
% \Procedure{path\_selection}{$circuit~K$}
%     \State largest\_cone $\leftarrow$ output port with largest cone;
%     \State b = BFS(largest\_cone);
%     \While {tail $\leftarrow$ pop(b)}
%         \If{ (Slack(tail) $>$ KeygateDelay and tail was not marked) }
%             \State path $\leftarrow$ DFS on tail considering slacks;
%             \State mark path as selected in the circuit;
%             \State\textbf{return} path;
%         \EndIf
%     \EndWhile
% \EndProcedure

% \end{algorithmic}
% \end{algorithm}

Our proposed solution for timing-aware cyclic obfuscation is presented in Algorithm \ref{switchinsertionalgo}. Both SC and LFN methods (supported in this algorithm) require selection of non-overlapping logic paths in the circuit for intertwined cycle creation. In our solution, presented in Algorithm \ref{switchinsertionalgo}, we find these non-overlapping logic paths in the fan-in cone (FIC) of a single primary output. The reason for selecting the logic paths in the same FIC is to take advantage of existing connections between selected logic sub-paths when one sub-path is in the FIC of at least one of the gates in the other sub-path. This condition results in the generation of many additional cycles, on top of those generated by LFN or SC. This is because each feedback could create a cycle when combined with each of path forward edges. After selection of a logic sub-path and before committing to the insertion of a new switch, the netlist is assessed for timing violation. If there is no violation, the cycle is generated and the slack of affected timing paths are updated. Finally, the logic gates in the selected sub-path are marked as \textbf{used}, removing them from future searches.

Our proposed algorithm selects new logic paths in the FIC of the selected primary output until there are no more viable sub-paths. The algorithm could be modified to continue finding new paths by selecting the next primary output candidate that has the largest number of un-\textbf{used} gates.

% This procedure is shown in Algorithm \ref{pathselectionalgo}. After determining the largest cone in the circuit using a simple backtracking on output ports, it will find the order of gates in that cone using a breadth-first search. Each unmarked gate in this list (in a timing path with enough slack for insertion of a multiplexer) could be considered as the tail of a sub path. Then, it will run a depth-first search on the selected tail to select a path for obfuscation.

% \begin{figure}[b]
% \centering
%     \includegraphics[width=0.6\columnwidth]{images/pathselection.pdf}
%     \caption{Selected paths in an output cone.}
%     \label{pathselection_example}
%     %  \vspace{-8pt}
% \end{figure}

\section{Results}\label{results}
In this section, we analyze the effectiveness of our proposed defense against SAT, CycSAT, and BeSAT attacks. For finding cycles in a netlist (after cyclic obfuscation), we implemented the cycle identification algorithm proposed in \cite{hawick2008enumerating} using C++. Considering that the source code for BeSAT was not openly available, we implemented the BeSAT attack based on the description in \cite{besat} using Yices SAT solver \cite{10.1007/978-3-319-08867-9_49}. Our computational platform is a Dell PowerEdge R620 equipped with Intel Xeon E5-2670 and 64GB of RAM. We used ISCAS-85 benchmarks listed and described in Table~\ref{iscas_table} to evaluate our solution and to compare it with the prior work. The timeout limit in our experiments is set to 10 hours: If an experiment does not conclude within the timeout limit, its table entry is marked as \say{t/o}. In an experiment, if the netlist is too small for insertion of the number of required feedbacks, its table entry is marked as \say{Netlist is Small (NiS)}.

%\textcolor{green}{During cyclic obfuscation, we experiment and increase the number of inserted feedbacks to show the rate of increase in the number of generated cycles. In small circuits, however, we can not insert the required number of feedbacks using the LFN or SC methodology. In this case, the associated table entry is marked as "Netlist is Small (NiS)". }

\begin{table}[t]
\footnotesize
\centering
\caption{Description of ISCAS-85 circuits used in this paper.}
\label{iscas_table}
\setlength\tabcolsep{1pt} % default value: 6pt
\setlength\extrarowheight{2pt}
\begin{tabular}{@{} cccc||cccc||cccc @{}}
\toprule 
Circuit  & \#Gates & \#PIs & \#POs & Circuit  & \#Gates & \#PIs & \#POs & Circuit  & \#Gates & \#PIs & \#POs \\ \midrule
c432  & 160     & 36  & 7  & c1355 & 546     & 41  & 32 & c3540 & 1669    & 50  & 22    \\
c499  & 202     & 41  & 32 & c1908 & 880     & 33  & 25 &  c5315 & 2307    & 178 & 123  \\
c880  & 383     & 60  & 26 & c2670 & 1269    & 233 & 140 & c7552 & 3513    & 207 & 108  \\
\bottomrule
\end{tabular}

% \vspace{-15pt}
\end{table}

\vspace{-10pt}
\subsection{Exponential Growth in The Number of Cycles}
\subsubsection{\textbf{Cyclification Using Super Cycles (SC)}}
The number of cycles created in ISCAS-85 benchmarks, when using N=1, 2, 3, 5, 10, and 15 MCs of size 7 (i.e., 7 gates in a cycle) for building a SC is reported in Table \ref{table_of_SC}. Using curve fitting techniques, the number of cycles in each netlist is also reported as a function of the number of feedbacks $X$, in form of $2^{mX}$, in which $m$ is the netlist-specific exponential acceleration factor. The minimum bound for $m$ (according to the discussion in section~\ref{supercyclesection}) when using SC is one. However, as reported in Table \ref{table_of_SC}, the value of $m$ is usually far larger than one, meaning there would be a far larger number of cycles than that expected from the SC-imposed minimum bound.

\begin{table}[t]
\footnotesize
\centering
\caption{The number of cycles reported during CycSAT attack. The exponential fitting function is in form of $c=2^{mX}$. }
\label{table_of_SC}
\setlength\tabcolsep{2.5pt} % default value: 6pt
\setlength\extrarowheight{2pt}
\begin{tabular}{@{} l *8c @{}}
\toprule
\multicolumn{1}{r}{Circuit} & N=1 & N=2 & N=3 & N=5 & N=10 & N=15 & m \\
\midrule
c432    & 3,384 & 23,879 & $4.6 * 10^5$ & NiS & NiS & NiS &  6.3 \\
c499    & 10    & 331    & 1528 & $1.4 * 10^6$ & NiS & NiS & 4.1 \\
c880    & 67    & 1,601  & 1,903 & $5.0 * 10^6$ & t/o & t/o &  4.5 \\
c1355   & 59    & 636    & $5.7 * 10^5$ & $1.9 * 10^9$ & t/o & t/o & 6.2 \\ 
c1908   & 13    & 294    & 12,594 & $1.3 * 10^7$ & t/o & t/o  &  4.8 \\
c2670   & 273   & 1,570  & 8,912 & $2.9 * 10^5$ & t/o & t/o &  3.6 \\ 
c3540   & 1,215 & 5,991  & $8.7 * 10^5$ & $4.9 * 10^8$ & t/o &  t/o  & 5.8 \\
c5315   & 162   & 4,869  & 6,650 & $1.2 * 10^9$ & t/o & t/o  &  6.0 \\
c7552   & 11    & 124    & 1,558 & $2.6 * 10^5$ & $1.2 * 10^9$ &  t/o  & 3.0 \\ \bottomrule
\end{tabular}
% \vspace{1ex}

% \raggedright{\textbf{t/o:} CycSAT does not conclude within the specified time limit. \\
% \textbf{NiS:} Circuit is too small for inserting the specified number of feedbacks.}
% \vspace{-10pt}
\end{table}

As illustrated in Table \ref{table_of_SC}, increasing the number of feedbacks exponentially increases the number of cycles, such that with only 15 feedbacks, the cycles in none of the netlists could be counted in 10-hour limit. Note that, the designer can exponentially increase CycSAT attack's pre-processing time, by linearly increasing the number of feedbacks. For executions resulted in timeout, we also confirmed that initiating CycSAT with incomplete \textit{NC} clauses traps the SAT solver in an infinite loop. Hence, the attacker can not complete the pre-processing in a reasonable time, and incomplete pre-processing traps the subsequent invocation of the SAT solver. The area overhead for building the SC in terms of the number of switches depends on the number of MCs and the number of gates in each MC. The area overhead for having various numbers of MCs of 7 gates when building a SC is reported in Table \ref{sc_area_overhead}. \\[-7pt]
\begin{table}[t]
% \footnotesize
\centering
\caption{Percentage of area overhead for SC creation when using  different number of MCs (N) of length 7.}
\label{sc_area_overhead}
\scalebox{0.95}{
\begin{tabular}{@{} *8c @{}}
\toprule 
\multirow{2}{*}{Circuit}  & N=1  & N=2   & N=3   & N=5   & N=10  & N=15  & N=20   \\
\cmidrule(lr){2-8}
 & \multicolumn{7}{c}{Area Overhead Percentages (\%)} \\
\midrule
c432  & 7.50 & 13.75 & 20.00 & NiS   & NiS   & NiS   & NiS   \\
c499  & 5.94 & 10.89 & 15.84 & 25.74 & NiS   & NiS   & NiS   \\
c880  & 3.13 & 5.74  & 8.36  & 13.58 & 26.63 & 39.69 & 52.74 \\
c1355 & 2.20 & 4.03  & 5.86  & 9.52  & 18.68 & 27.84 & 37.00 \\
c1908 & 1.36 & 2.50  & 3.64  & 5.91  & 11.59 & 17.27 & 22.95 \\
c2670 & 0.95 & 1.73  & 2.52  & 4.10  & 8.04  & 11.98 & 15.92 \\
c3540 & 0.72 & 1.32  & 1.92  & 3.12  & 6.11  & 9.11  & 12.10 \\
c5315 & 0.52 & 0.95  & 1.39  & 2.25  & 4.42  & 6.59  & 8.76  \\
c7552 & 0.34 & 0.63  & 0.91  & 1.48  & 2.90  & 4.33  & 5.75  \\ \bottomrule
\end{tabular}
}
% \vspace{1ex}

% \raggedright{\textbf{NiS:} Circuit is too small for inserting the specified number of feedbacks.}
% \vspace{-10pt}
\end{table}

\subsubsection{\textbf{Cyclification using Logarithmic Feedback Networks (LFN)}}
As discussed and proved in section \ref{buildingLFN}, the lower bound on the number of generated cycles, when the LFN method for cyclification flow is adopted, is an exponential function of the number of feedbacks. Furthermore, similar to SC, the edge density of the original netlist may substantially increase the number of created cycles. This is because of the gates with fan-outs greater than one in selected logic path segments. If the output of a gate in the LFN is connected to the input(s) of another gate(s) in the same network, the resulting net counts as an additional forward path. Then, each forward path could be matched with a feedback, resulting in an additional cycle. Considering that path segments are selected from the FIC of the same primary output, there exist many such connections (forward edges), resulting in the generation of a far larger number of cycles than the guaranteed minimum bound expected from using LFN. To illustrate this, both the number of created cycles for each benchmark and the theoretical lower bound (calculated using $\sum_{l=1}^N {N \choose l}(l - 1)!$ as proved in section \ref{buildingLFN}) is reported in Table \ref{lfn_table}. As illustrated, for most of the obfuscated benchmarks with a LFN larger than 4, cycle enumeration results in timeout after 10 hours due to the exponential number of created cycles. This indicates an exponential runtime at the CycSAT pre-processing stage. The area overhead for creating LFNs of different sizes (different number of input paths) is reported in Table \ref{lfn_area_overhead}. Note that in both SC and LFN, the area overhead scales with the number of inserted feedbacks and not the size of the circuit. Hence, the area overhead is smaller in larger circuits.

\begin{table}[t]
\scriptsize
\centering
\caption{Number of cycles reported during CycSAT attack using LFN method. N is the number of selected paths for creating LFN.}
\label{lfn_table}
\setlength\tabcolsep{2.5pt} % default value: 6pt
\setlength\extrarowheight{2pt}

\begin{tabular}{@{} cccccc @{}}
\toprule 
% lower bound calculated using
% http://www.wolframalpha.com/input/?i=sum(l%3D1,l%3Dm)(combination(m,l)*((l-1)!))+where+m%3D32
 & N=2 & N=4 & N=8 & N=16 & N=32 \\ \midrule
Lower Bound & 3 & 24 & 16072 & $3.8 \times 10^{12}$ & $2.3 \times 10^{32}$ \\ \midrule
c432 & 26,578 & NiS & NiS & NiS & NiS \\
c499 & 192 & 278,577 & $1.3 \times 10^{10}$ & t/o & NiS \\
c880 & 8,836 & $4.5 \times 10^8$ & t/o & t/o & NiS \\
c1355 & $8.3 \times 10^6$ & t/o & t/o & t/o & NiS \\
c1908 & $8.4 \times 10^7$ & t/o & t/o & t/o & t/o \\
c2670 & $1.2 \times 10^7$ & t/o & t/o & t/o & t/o \\
c3540 & $8.5 \times 10^9$ & t/o & t/o & t/o & t/o \\
c5315 & $1.2 \times 10^9$ & t/o & t/o & t/o & t/o \\
c7552 & $2.9 \times 10^9$ & t/o & t/o & t/o & t/o \\
\bottomrule
\end{tabular}

% \vspace{1ex}

% \raggedright{\textbf{t/o:} CycSAT pre-processing does not conclude within the specified time limit. \\
% \textbf{NiS:} Circuit is too small for inserting the specified number of feedbacks.}
\end{table}

Capturing the power overhead of cyclic obfuscation is more involved. The leakage component of power overhead is a function of the area overhead of the obfuscation solution, and threshold voltage (VT) of inserted multiplexers. Using a high-VT switch cell reduces the leakage impact, however, it introduces additional delay \cite{irataVakil}. In a simple implementation where standard cells are selected from a single VT, the increase in the leakage power is similar to the increase in the area. The dynamic power consumption, on the other hand, depends on the switching activity of the inserted switches. After proper activation, the switching activity of the inserted multiplexers depends on the toggling rate of the correct input net to the multiplexer. The net toggling activity, in turn, depends on the level of controllability of that net and the probable input scenario to the netlist. The power consumption of both the LFN and SC-based solutions of size N=16 is provided in Table~\ref{power_table}. However, note that the power consumption could improve (at the expense of timing and security) by modifying the SC or LFN algorithm to choose nets with small toggling rate to reduce the overhead of dynamic power consumption.

\begin{table}
% \footnotesize
\centering
\caption{Percentage of area overhead for an inserted LFN for different number of selected paths (N).}
\label{lfn_area_overhead}
\setlength\extrarowheight{2pt}
\scalebox{0.9}{
\begin{tabular}{@{} *6c @{}}
\toprule 
\multirow{2}{*}{Circuit}  & N=2  & N=4   & N=8   & N=16   & N=32   \\
\cmidrule(lr){2-6}
 & \multicolumn{5}{c}{Area Overhead Percentages (\%)} \\
\midrule
c432  & 5.00 & NiS & NiS & NiS & NiS \\
c499  & 3.96 & 11.88 & 27.72 & 79.21  & NiS    \\
c880  & 2.09 & 6.27  & 14.62 & 41.78  & NiS    \\
c1355 & 1.47 & 4.40  & 10.26 & 29.30  & NiS    \\
c1908 & 0.91 & 2.73  & 6.36  & 18.18  & 43.64  \\
c2670 & 0.63 & 1.89  & 4.41  & 12.61  & 30.26  \\
c3540 & 0.48 & 1.44  & 3.36  & 9.59   & 23.01  \\
c5315 & 0.35 & 1.04  & 2.43  & 6.94   & 16.64  \\
c7552 & 0.23 & 0.68  & 1.59  & 4.55   & 10.93  \\ \bottomrule
\end{tabular}
}
% \vspace{1ex}

% \raggedright{\textbf{NiS:} Circuit is too small for inserting the specified number of feedbacks.}
\end{table}

\begin{table}[t]
% \scriptsize
\centering
\caption{The power overhead of SC and LFN of size N=16.}
\label{power_table}
\scalebox{0.95}{
\begin{tabular}{@{} ccccc @{}}
\toprule 
\multirow{2}{*}{Circuit}  & \multicolumn{2}{c}{SC (N=16)}  & \multicolumn{2}{c}{LFN (N=16)}   \\
\cmidrule(lr){2-3}
\cmidrule(lr){4-5}
 & Switching (\%) & Leakage (\%) & Switching (\%) & Leakage (\%) \\
\midrule
c432  & NiS & NiS & NiS & NiS \\
c499  & NiS & NiS & 212.64 & 75.13   \\
c880  & 38.09 & 44.85  & 56.67 & 38.82   \\
c1355 & 12.79 & 32.77  & 13.26 & 24.6   \\
c1908 & 8.42 & 19.1  & 13.38  & 15.66   \\
c2670 & 14.14 & 15.96  & 13.17  & 12.32   \\
c3540 & 8.76 & 10.79  & 3.86  & 8.88    \\
c5315 & 5.75 & 8.51  & 6.13  & 6.7    \\
c7552 & 2.88 & 5.78  & 7.79  & 4.4    \\ \bottomrule
\end{tabular}
}
% \vspace{1ex}

% \raggedright{\textbf{NiS:} Circuit is too small for inserting the specified number of feedbacks.}
\end{table}

\begin{table*}
\scriptsize 
\centering
\caption{SAT attack, CycSAT, and BeSAT execution time after insertion of a SC (N=2), insertion of a SC and 10 SR-latches (N=2 + SR-L=10), and insertion of 15 MCs and 10 SR-latches (N=15 + SR-L=10).}
\label{sat_results}
\setlength\extrarowheight{2pt}
\begin{tabular}{@{} cccccccccccccc @{}}
\toprule 
\multirow{2}{*}{Circuit} & \multicolumn{3}{c}{N=2} & \multicolumn{5}{c}{N=2 + SR-L=10} & \multicolumn{5}{c}{N=15 + SR-L=10} \\ 
\cmidrule(lr){2-4} \cmidrule(lr){5-9} \cmidrule(lr){10-14}
 & SAT & \#Cycles & CycSAT-I & SAT & \#Cycles & CycSAT-I & CycSAT-II & BeSAT & SAT & \#Cycles & CycSAT-I & CycSAT-II & BeSAT
\\ \midrule
c432    & Inf  & 23,879 & 2.56s  & Inf & $1.65 \times 10^5$ & UNSAT & 11.69s & 35.48s & Inf & t/o & UNSAT & t/o & t/o \\ 
c499    & 0.56s & 236 & 0.10s  & Inf & 397 & UNSAT & 0.11s & 0.79s & Inf & t/o & UNSAT & t/o & t/o \\
c880    & Inf  & 1,601 & 0.24s  & Inf & $7.87 \times 10^6$ & UNSAT & 793.12s & t/o & Inf & t/o & UNSAT & t/o & t/o \\
c1355   & Inf  & 636 & 0.12s  & Inf & $5.00 \times 10^5$ & UNSAT & 53.21s & 134.56s & Inf & t/o & UNSAT & t/o & t/o \\
c1908   & 0.28s & 294 & 0.10s  & Inf & 6,467 & UNSAT & 0.73s & 170.74s & Inf & t/o & UNSAT& t/o & t/o \\
c2670   & Inf  & 1,570 & 0.23s  & Inf & 7,412 & UNSAT & 0.92s & 17.22s & Inf & t/o & UNSAT & t/o & t/o \\
c3540   & Inf  & 5,991 & 0.75s  & Inf & 6,026 & UNSAT & 0.75s & 22.67s & Inf & t/o & UNSAT & t/o & t/o \\
c5315   & Inf  & 4,869 & 0.61s  & Inf & $2.59 \times 10^5$ & UNSAT & 26.04s & 370.08s & Inf & t/o & UNSAT & t/o & t/o \\
c7552   & Inf  & 124 & 0.189s  & Inf & 164 & UNSAT & 0.19s & 18.30s & Inf & t/o & UNSAT & t/o & t/o \\ \bottomrule
\end{tabular}
% \vspace{1ex}

% \raggedright{\textbf{t/o:} Attack does not conclude within the specified time limit. \\
% \textbf{Inf:} SAT solver enters an infinite loop.}
% \vspace{-4ex}
\end{table*}

% \vspace{-10pt}
\subsection{SAT, CycSAT and BeSAT Attack Resilience}
Table \ref{sat_results} captures the result of SAT, CycSAT, and BeSAT attacks on ISCAS-85 benchmarks that are obfuscated using our proposed solution. For generating the data in this table, we prepared three sets of obfuscated benchmarks. The first set of benchmarks is obfuscated with only two MCs using the SC approach for obfuscation method. This group of obfuscated benchmarks represents cyclification with a small number of dummy cycles, with no real cycles. The netlists in the second set, are first obfuscated using 10 SR-latches (by using the input-dependency based obfuscation as described in section \ref{idbo}) and then are cyclified by inserting two MCs. The second group represents the case where there are some real cycles in the design, while the total number of cycles is still small. The third group is similar to the second group, however, the number of inserted MCs is increased to 15. It represents obfuscated solutions with both real and exponentially large number of dummy cycles. The results of running SAT, CycSAT, and BeSAT is captured in Table \ref{sat_results}. For c432 and c499, generating large number of MCs (15) was not possible, hence, the largest number of possible MCs were used in the generation of SC.

The first group introduces a small number of removable cycles. As reported in Table \ref{sat_results}, even the existence of simple cycles traps the original SAT attack in an infinite loop in most cases (except for two benchmarks that SAT solver luckily chooses a sequence of inputs that avoid or exit the trap). However, CycSAT, when uses the "no structural path" condition (CycSAT-I) for generating the cycle avoidance clauses, easily breaks all obfuscated netlists. As illustrated in this table and predicted in Equation \ref{eq:preproceTime}, CycSAT runtime (which includes the runtime for both pre-processing step and SAT solver's invocation) almost linearly varies with the number of cycles in each netlist.

For the second group, where the original circuit is also cyclified (using real cycles), the usage of CycSAT-I returns UNSAT as it produces \textit{NC} clauses that breaks the real Boolean cycles. However, when CycSAT uses the "no sensitizable path" conditions (CycSAT-II), it breaks the obfuscation in all cases. Most notable in this data is the increase in the runtime of CycSAT attack (when compared to the first group) as the time it takes to compose the \textit{NC} condition for each cycle based on "no sensitizable path" condition is longer. This validates the impact of logic cyclification on the runtime of CycSAT attack. Another attack possibility is BeSAT attack. However, the BeSAT attack should be slightly modified: considering that the design contains real Boolean cycles, the \say{no sensitizable path} condition (instead of \say{no structural path} in the BeSAT attack as described in \cite{besat}) should be used for the generation of the \textit{NC} clauses. Hence, the attack could be carried by generating a set of \textit{NC} clauses (given a deadline) and then use BeSAT to attack the obfuscation and recover from oscillating and stateful cycle conditions. To model this attack, we set \say{no sensitizable path} pre-processing deadline to 2 hours, and BeSAT attack time to 8 hours (total of 10 hours attack time). As shown for \say{N=2+SR-L=10}, all but one deobfuscation was successful and in general, BeSAT underperform compared to CycSAT-II attack. This is because there exists a small number of cycles, and both CycSAT-II and BeSAT have found and conditioned all cycles, however, BeSAT due to the runtime monitoring of DIPs is slower compared to CycSAT-II attack.

Finally, for the third group, where the number of inserted feedbacks is increased to 15, all three attacks fail. The CycSAT-I is not applicable, as it will open real cycles, resulting in netlist malfunction, and even if pre-processing of this attack finishes it will exit as UNSAT. The CycSAT-II fails as it can not finish the pre-processing on time. Note that by increasing the number of feedbacks, the designer can easily and exponentially increase the required pre-processing time unreasonably long. The remaining attack possibility is the BeSAT attack. In this case, the pre-processing of \textit{NC} clauses is carried until the time limit (2 hours) and then BeSAT attack is carried out. Note that in this condition, the BeSAT starts the SAT attack with a partial set of clauses generated in the pre-processing step. However, as illustrated in Table \ref{sat_results}, BeSAT will reach the deadline after invalidating 100s of thousands of keys. This is when there exist millions (or larger) other keys that cause oscillating behavior which BeSAT has not yet examined and pruned (one at a time) in the time limit.

As explained in section IV, BeSAT only works when the number of undetected cycles (and un-conditioned keys) is small. The BeSAT attack is slow and eliminates one-incorrect-key at a time. This is when, in our proposed obfuscation solution, there exists an exponentially large number of invalid keys even after partial pre-processing: As a part of our obfuscation solution (and to create real cycles), we are using (diffused) SR-latches. To prevent stateful behavior, through careful input-logic section (as described in section \ref{idbo}), we ensure that the value of ‘$SR$’ input can not evaluate to ‘$11$’ (condition for statefulness). For this purpose, the input logic cone to $S$ and $R$ input is constructed by exploiting the interdependency of selected wires in the netlist. However, the selection of inputs is further hidden through routing obfuscation. In this case, only with the application of the correct key, the interdependence of the input wires will render the SR-latch non-stateful (by skipping the 11 input). Let's assume $S = g(K_1, X)$ and $R = f(K_2, X)$, where the $g$ and $f$ are the logic representing the input cone of $S$ and $R$ input to the SR-latch, $K_1$ and $K_2$ are the key gates in the fan-in cone of $S$ and $R$, and the $X$ is the choice of primary input. In this scenario, any choice of $K_1$, $K_2$ and $X$ that could make the $SR=11$ will result in a stateful circuit. From this analysis, the worst-case scenario for BeSAT is a function of the size of primary input $X$, and key selection $K_1$ and $K_2$ for which the wire $S$ and $R$ evaluate to $1$, which is an exponential function of the key-length $K = (K_1 \bigcap K_2)$. Considering that our solution builds a strongly-connected graph, the FIC of $S$ and $R$ could span to all the key-gates. Hence, the number of invalid keys that should be banned is exponentially large. Considering this discussion, and for a large number of key combinations that should be banned (one at a time), as shown in the results for \say{N=15+SR-L=10}, BeSAT attack does not work against our proposed solution.

% For the sake of completeness of the study, it is necessary to address one other attack methods that were discussed in section \ref{cycobfuscation}. In the attack proposed by Chen \cite{Chen:2018:ESA:3217208.3190853}, only the dummy cycles could be detected and removed in the circuits. In our proposed obfuscation method, both dummy and real cycles are added to the circuits. Hence, this attack, by opening the real cycles (inserted using the three methods proposed in section IV.B) will render the circuit non-functional. Hence, this attack would fail in extracting the correct key in all benchmarks obfuscated using our proposed solution, and for that reason is not reported in Table \ref{sat_results}.

\subsection{SAT, CycSAT and BeSAT Resiliency of Previous Methods}
In this section, we study the effectiveness of previously proposed cyclic logic solutions and compare them with our proposed solution. To attack the prior art solutions we use the modified CycSAT attack as described and formulated in section \ref{breaking_hard_cycles} of this paper. The modified CycSAT attack works similar to the original CycSAT attack, however, instead of composing the \textit{NC} clauses per detected feedback, it composes the \textit{NC} clauses per detected cycle.

% For a comprehensive study of cyclic logic locking and for evaluating the strength of modified CycSAT attack as proposed in section \ref{breaking_hard_cycles} of this paper (by generating "no structural path" for all cycle), it is useful to compare the proposed attack methods with the previous obfuscation schemes. 

The original cyclic locking method was introduced in \cite{Shamsi:2017:COC:3060403.3060458} where authors proposed inserting multiplexers in the circuit to create cycles. This obfuscation solution attempts to create irreducible cycles. This method can only create dummy cycles as it does not affect the DAG nature of a combinational netlist and is referenced in this paper as \textit{glsvlsi17}. The second method discussed here \cite{ZhouCycles} considers CycSAT attack and tries to defeat CycSAT-I using an auxiliary-circuit. This method was discussed in section \ref{nested_cycles}. By adding the proposed auxiliary-circuits to a design, real cycles are formed, converting the DAG nature of the netlist to a DCG. The netlist is then augmented with additional dummy cycles (similar to the glsvlsi17 method), making the netlist to contain both real and dummy cycles. In this paper, we use the name \textit{date18} to refer to this cyclic obfuscation solution.

To assess the effectiveness of prior art solutions, we modeled each of the glsvlsi17 and date18 to obfuscate the ISCAS-85 benchmarks. To compare the evaluation results of prior art to that of our proposed solution (in Table \ref{sat_results}), the glsvlsi17 method is implemented using 15 randomly selected feedbacks of length 7, while the benchmarks prepared using date18 solution are obfuscated using the same number of feedbacks (15) and 10 real cycles (for DAG to DCG transformation), implemented using the auxiliary-circuit as described in \cite{ZhouCycles}. For smaller benchmarks, where insertion of this many feedbacks was not feasible, we have inserted the largest feasible number of feedbacks. To show the effectiveness of our solution in increasing the runtime of the CycSAT pre-processing step, we have also evaluated the number of generated cycles for each of the prior cyclic obfuscation (glsvlsi17 and date18) solutions.

Table \ref{glsvlsi17_res} captures our evaluation results for glsvlsi17 when attacked using SAT, CycSAT-I, and BeSAT. As expected the success of SAT attack on selected benchmarks is random, as generated cycles could trap the SAT solver. Note that by increasing the number of feedbacks, the chances of trapping the SAT solver increases. CycSAT-I breaks the obfuscation and finds the key to all but one obfuscated benchmark. For c1355, cycles could not be processed within the 10-hour time limit, and the attack is timed out. But this case is a great showcase to see the power of BeSAT. As expected, BeSAT could also break this obfuscation. Considering that the pre-processing for most of the benchmarks could be done in less than 2-hours, and all cycles could be found for such small obfuscations, the number of banned keys for all cases but one is zero. For this reason and for the additional overhead of runtime monitoring of SAT execution time, the BeSAT takes longer than CycSAT-I. The only interesting scenario is for c1355, where the CycSAT-I is timed out and can not finish the pre-processing of all cycles. In this case, the incomplete set of \textit{NC}s is used in BeSAT, and with only 3 banned keys, BeSAT skips the traps and finds the correct key. Note that the reason why BeSAT does work is that the number of oscillating keys generated in this obfuscation solution is small. This is unlike our proposed solution that there exists an exponentially large number of such keys, and if given to BeSAT, they have to be eliminated one at a time.

% \textcolor{blue}{By considering the key-banning process in BeSAT, this attack could recover the correct key for c1355 in 20.83s after generating "no structural path" conditions for two hours. Then, SAT solver ran into three stateful keys during DIP generation process that could stuck the CycSAT-I in an infinite loop but in here, the attack could exit with the correct key.}

\begin{table}[t]
% \scriptsize
\centering
\caption{SAT attack, modified CycSAT, and BeSAT results for evaluation of glsvlsi17 method \cite{Shamsi:2017:COC:3060403.3060458}.}
\label{glsvlsi17_res}
\setlength\tabcolsep{3pt} % default value: 6pt
\begin{tabular}{@{} cccccccc @{}}
%c|c|cc|cc
\toprule 
\multirow{ 2}{*}{Circuit} & \multirow{ 2}{*}{\#Cycles} & \multicolumn{2}{c}{SAT} & \multicolumn{2}{c}{CycSAT-I} & \multicolumn{2}{c}{BeSAT} \\ 
\cmidrule(lr){3-4}
\cmidrule(lr){5-6}
\cmidrule(lr){7-8}
 &  & Time & Iteration & Time & Iteration & Time & Banned \\
\midrule
c432  & 32 & t/o & - & 0.02 & 1 & 0.45 & 0 \\
c499  & 282 & t/o & - & 0.05 & 1 & 0.88 & 0 \\
c880  & 36 & 1.35  & 61  & 0.13 & 15 & 3.59 & 0 \\
c1355 & t/o & t/o & - & t/o & - & 7220.83 & 3 \\
c1908 & 1,625 & t/o & - & 0.95 & 83 & 8.44 & 0 \\
c2670 & 129 & t/o  & -  & 2.26  & 19 &  55.11 & 0 \\
c3540 & 606 & 0.63 & 41 & 0.70 & 14 & 10.03 & 0 \\
c5315 & 4,216 & 1.7 & 33 & 1.19  & 45 & 32.75 & 0 \\
c7552 & 1,117 & 2.35 & 105 & 1.77  & 73 & 43.24 & 0 \\ \bottomrule
\end{tabular}
% \vspace{1ex}

% \raggedright{\textbf{t/o:} Attack execution does not conclude within the specified time limit.}
\end{table}

\begin{table}
\scriptsize
\centering
\caption{Evaluating date18 obfuscation \cite{ZhouCycles} against SAT, CycSAT and BeSAT.}
\label{date18_res}
\setlength\tabcolsep{2.5pt} % default value: 6pt
\begin{tabular}{@{} cccccccccc @{}}
\toprule 
\multirow{ 2}{*}{Circuit} & \multirow{ 2}{*}{\#Cycles} & \multicolumn{2}{c}{SAT} & \multicolumn{2}{c}{CycSAT-I} & \multicolumn{2}{c}{CycSAT-II} & \multicolumn{2}{c}{BeSAT} \\ 
\cmidrule(lr){3-4}
\cmidrule(lr){5-6}
\cmidrule(lr){7-8}
\cmidrule(lr){9-10}
 &  & Time & Iteration & Time & Iteration & Time & Iteration & Time & Banned  \\
\midrule
c432  & 62 & t/o & - & 0.02 & UNSAT & 0.3 & 18 & 9.19 & 0 \\
c499  & 1,157 & t/o & - & 0.06 & UNSAT & 0.14 & 18 & 2.00 & 0  \\
c880  & 56 & t/o  & -  & 0.04 & UNSAT & 0.31 & 23 & 5.11 & 0  \\
c1355 & t/o & t/o & - & t/o & - & t/o & - & 7268.77 & 12 \\
c1908 & 1,645 & 1.99 & 144 & 0.02 & UNSAT & 0.88 & 68 & 205.02 & 0 \\
c2670 & 149 & t/o  & -  & 0.03  & UNSAT & 0.53 & 41 & 10.53 & 0 \\
c3540 & 626 & 6.49 & 187 & 0.1 & UNSAT & 1.53 & 37 & 18.19 & 0 \\
c5315 & 4,236 & t/o  & - & 0.05 & UNSAT & 2.06 & 60 & 30.45 & 0 \\
c7552 & 1,137 & t/o  & -  & 0.08 & UNSAT & 1.9 & 31 & 40.75 & 0 \\ \bottomrule
\end{tabular}
% \vspace{1ex}

% \raggedright{\textbf{t/o:} Attack execution does not conclude within the specified time limit.}
% \vspace{-4ex}
\end{table}

% By modifying the signal selection algorithm in glsvlsi17 from a random to a targeted one, it is possible to create a situation similar to the c1355 for other circuits. If feedbacks, instead of being connected to a random node, are connected to a node in the FIC of  cones, all random signals also result in cycles. It is shown in Fig. \ref{exp_cycles_mux} for an example circuit. Two types of random signals are needed for this obfuscation method, feed-out signal and feed-in signal to/from the middle wire. Feed-out connection point should be selected from the fan-in cone and feed-in signals should be selected from fan-out cone of a wire.

Table \ref{date18_res} captures evaluation results for date18 method. Aware of the shortcomings of glsvlsi17, the date18 solution was proposed as a CycSAT-resistant obfuscation solution. The proposed auxiliary-circuit by itself has a minimal impact on the number of cycles. However, this method is expected to have a larger number of stateful cycles, and when the original SAT attack used there are higher chances for trapping the SAT solver in an infinite loop. The results in Table \ref{date18_res} support this hypothesis, as only two benchmarks are successfully attacked using the base SAT attack. When attacked using CycSAT-I, the date18 solution remains resistant as the pre-processing step of CycSAT-I incorrectly opens the real cycles during \textit{NC} clause generation. However, when the modified CycSAT-II attack, as described in section \ref{breakingCycSAt}, is deployed, could easily break all instances of obfuscated solutions except c1355 (that could not be pre-processed in a reasonable time for having a very large number of cycles). However, in the case of BeSAT and after limiting the pre-processing time to two hours, the key for c1355 could be recovered in 68.77s after 2 hours of \textit{NC} clause generation. Other benchmarks that previously was broken by CycSAT-II are also broken by BeSAT with zero banned keys since the generated \textit{NC} clauses cover all undesirable cycle conditions. Note that for this attack, the \textit{NC} clauses for BeSAT are generated using the \say{no sensitizable path} condition, otherwise the attack will return as UNSAT.

Comparing the glsvlsi17 and date18 data in Tables \ref{glsvlsi17_res} and \ref{date18_res} with that of our proposed solution in Table \ref{sat_results} illustrate the effectiveness of our solution: none of the obfuscated netlists using our solution could be broken by SAT, CycSAT-I, CycSAT-II, or BeSAT (original and modified) attacks, as it includes a solution to trap both the SAT solver and pre-processing step of CycSAT/BeSAT. Note that, when deploying SAT or CycSAT attack to break glsvlsi17 or date18, the runtime, in addition to the number of inserted feedbacks, also depends on the selection of feedbacks. Hence, a random selection of feedbacks in glsvlsi17 and date18 results in considerable variation in the attack time. Therefore, these solutions, unlike our proposed solution, can not guarantee a monotonic increase in the runtime of the attack as the number of randomly selected feedbacks increases. Note that in our solution, the runtime is dominated by CycSAT's or BeSAT's pre-processing step, and this runtime is linearly dependent on the number of cycles, and the number of cycles is an exponential function of the number of inserted feedbacks. Hence, we can guarantee a monotonic increase in the overall runtime of the attack against our proposed solution as the number of inserted feedbacks increases.

% Using this technique the CycSAT-I could be defeated. CycSAT-I tries to break all the cycles that are introduced using multiplexers, and by opening real cycles, it will create an incorrect behavior in the design. This method further obfuscates the circuit by introducing hard cycles to defeat CycSAT-II. Although, this type of cycle could be missed in the original CycSAT attack, it would found in here due to the proposed modifications in section \ref{breakingCycSAt} to the pre-processing step. This method is mentioned as date18 in the results.

% \begin{figure}
%     \centering 

%     \begin{subfigure}[b]{\columnwidth}
%     \includegraphics[width=\columnwidth]{images/glsvlsi1.pdf}
%     \caption{}
%     % \vspace{5pt}
%     \end{subfigure}

%     \begin{subfigure}[b]{\columnwidth}
%     \includegraphics[width=\columnwidth]{images/glsvlsi2.pdf}
%     \caption{}
%     % \vspace{5pt}
%     \end{subfigure}

%     \caption{Modifying random signal selection for glsvlsi17 method. The random signals should be selected from fan-in and fan-out cones.}
%     \label{exp_cycles_mux}
% \end{figure}

\vspace{-10pt}
\subsection{Timing Aware Cyclification}
As described in section \ref{timawareobf}, inserting logic gates in timing-critical paths would increase the critical path of the netlist resulting in a performance penalty. To minimize the performance penalty to the extent possible, we proposed a timing aware cyclic obfuscation flow in section \ref{timawareobf}. This solution would only affect the timing if it can no longer use non-critical timing paths for feedback insertion.  

Table \ref{table_of_ta} captures the result of our proposed timing aware cyclic obfuscation when allowing 0\% and 5\% delay overhead for cyclic obfuscation. Using this delay constraint, the algorithm tries to insert the maximum number of feasible feedbacks in each benchmark using the SC solution proposed in section \ref{supercyclesection}. In this table, we have provided a measure of the maximum number of MCs that could be implemented in each benchmark for building a strongly connected graph before running out of usable gates. The key count is the sum of the number of key values needed for managing the MCs and the number of key values needed for managing the additional multiplexers (used for creating outgoing edges from internal gates in each MC). As illustrated, the maximum number of MCs and key values is a function of the netlist size and the acceptable delay overhead. Note that in larger benchmarks, even without incurring a time penalty we can insert a large number of MCs, pushing CycSAT attack to be trapped in its pre-processing step until timeout. In addition, note that with 10 MCs, our C++ implementation of pre-processor can not finish counting the number of generated cycles, and according to SC and LFN lemmas proved in sections \ref{supercyclesection} and \ref{buildingLFN}, the number of generated cycles exponentially grows with each added feedback. Hence, we can make the attack-time unreasonably long with no or limited timing impact.   

\begin{table}[t]
% \scriptsize
\footnotesize
\centering
\caption{Timing-aware obfuscation results for the Super Cycle method. Maximum number of Micro Cycles are inserted for 0\% and 5\% overhead over timing slack.}
\label{table_of_ta}
\setlength\tabcolsep{1.25pt} % default value: 6pt
\setlength\extrarowheight{2pt}
\scalebox{0.9}{
\begin{tabular}{@{} ccccc|ccccc @{}}
\toprule
\multirow{ 2}{*}{Circuit} & \multicolumn{4}{c}{Slack = 5\%} & \multicolumn{5}{c}{Slack = 0\%}   \\
\cmidrule(lr){2-5}
\cmidrule(lr){6-10}
      & \#Cycles    & SAT(s)    & \#Keys  & \#MCs   & \#Cycles & SAT(s) & \#Keys & \#MCs & Area \% \\
\midrule      
c432  & 303,476 & 0.14     & 15   & 2   & NiS        & NiS    & NiS  & NiS     & NiS      \\
c499  & NiS    & NiS  & NiS & NiS & NiS     & NiS         & NiS       & NiS    & NiS     \\
c880  & t/o    & t/o  & 95  & 18    & t/o       & 0.23      & 51 & 12    & 26.63   \\
c1355 & t/o    & t/o  & 109 & 23    & 2,766      & 0.55      & 25 & 8    & 9.16    \\
c1908 & t/o    & t/o  & 187 & 38    & t/o       & t/o   & 111 & 24   & 25.23   \\
c2670 & t/o    & t/o  & 335 & 70    & t/o       & t/o   & 244 & 53   & 38.46   \\
c3540 & t/o    & t/o  & 378 & 75    & t/o       & t/o   & 274 & 57   & 32.83   \\
c5315 & t/o    & t/o  & 448 & 110    & t/o       & t/o   & 446 & 95   & 38.66   \\
c7552 & t/o    & t/o  & 729  & 183   & t/o       & t/o   & 632 & 158   & 35.98   \\
\bottomrule
\end{tabular}
}
% \vspace{1ex}

% \raggedright{\textbf{t/o:} Attack execution does not conclude within the specified time limit. \\
% \textbf{NiS:} Circuit is too small or available slack prevents inserting the specified number of feedbacks.}
% \vspace{-4ex}
\end{table}

% deploying an attack against obfuscated benchmarks larger than c880, the number of possible feedbacks are enough to insert more than 10 MCs. This is enough to create a SAT- and CycSAT-resilient circuit. It means that creating exponential number of cycles is feasible in most cases without introducing any additional delays on the circuit and the only limiting factor is the area overhead.

The number of MCs and the number of gates in each MC (e.g., cycle length) could affect the number of created cycles and defines the SAT resiliency of the circuit. Parameters like targeted frequency and area overheads should also be considered during cyclic obfuscation. However, this could create a trade-off on how SAT-resilient a circuit is versus how efficiently it could be implemented.

% without introducing additional delay, we can introduce this many cycles and hence the additional security does not always come with the cost of additional area. This is because although a design may be Pareto optimal in 3 dimension of (PPA), it may not be Pareto optimal in space (SPPA). This will make it possible to get some security coverage at no PPA cost, by pushing the design to a SPPA optimal point. But after the design is SPPA Pareto optimal, the SPPA metrics should be traded off for one another.

%\vspace{-10pt}
\section{Conclusion}\label{conclusion}
In this paper, we proposed a new mean of cyclic obfuscation that is immune to SAT, CycSAT and BeSAT attacks. To make the pre-processing step of CycSAT and BeSAT attacks ineffective, we proposed two mechanisms (SC and LFN) for exponentially increasing the number of generated cycles with respect to the number of inserted feedbacks. In addition, we proposed three mechanisms to cyclify the circuit with real cycles (Cyclic Boolean Logic). The addition of real cycles forces an attacker to generate the \say{no sensitizable path} conditions during the pre-processing step of CycSAT or BeSAT attacks, which is considerably more time consuming than \say{no structural path} generation. The exponential increase in the number of feedbacks prevents the attacker from generating \textit{NC} conditions for all cycles in a reasonable amount of time. This breaks CycSAT attack. The BeSAT attack can proceed to its SAT stage with an incomplete set of \textit{NC} clauses, however, it has to ban remaining invalid keys one at a time, and there exists an exponentially large number of such keys. Hence, it also fails to break the proposed solution. 
 
%  To create exponentially large number of dummy cycles, we proposed two new techniques. introduced several techniques for exponentially increasing the number of cycles with respect to the number of inserted feedbacks. This, in turn, resulted in an exponential increase in runtime of the pre-processing step, preventing the SAT attack completion. Based on this study, the cyclic obfuscation when properly implemented poses an exponential runtime on CycSAT attack with respect to the number of inserted feedbacks. Hence, CycSAT or existing SAT-attacks are not an effective means for breaking cyclic obfuscation.

% By introducing cycles in a netlist, the straightforward SAT-attack would be trapped in an infinite loop. However, an attacker could use CycSAT or BeSAT attacks to break such obfuscation problems. The problem with both CycSAT and BeSAT is the runtime of their pre-processing step for generating the cycle avoidance clauses, which grow as a linear function of the number of cycles in the netlist. As a means of defense, we introduced several techniques for exponentially increasing the number of cycles with respect to the number of inserted feedbacks. This, in turn, resulted in an exponential increase in runtime of the pre-processing step, preventing the SAT attack completion. Based on this study, the cyclic obfuscation when properly implemented poses an exponential runtime on CycSAT attack with respect to the number of inserted feedbacks. Hence, CycSAT or existing SAT-attacks are not an effective means for breaking cyclic obfuscation.

\section*{Acknowledgement} \label{Acknowledgement}
This research is funded by the Defense Advanced Research Projects Agency (DARPA, \#FA8650-18-1-7819) of the USA, and joint funding from National Science Foundation and Semiconductor Research Corporation (Award \#1718434). \\ 
\vspace{-2.5ex}
\renewcommand{\IEEEbibitemsep}{0pt plus 0.5pt}
\makeatletter
\IEEEtriggercmd{\reset@font\normalfont\fontsize{7.0pt}{8pt}\selectfont}
\makeatother
\IEEEtriggeratref{1}

\bibliographystyle{IEEEtran}
\bibliography{IEEEabrv,refs}

% Generated by IEEEtran.bst, version: 1.14 (2015/08/26)
\begin{thebibliography}{10}
\providecommand{\url}[1]{#1}
\csname url@samestyle\endcsname
\providecommand{\newblock}{\relax}
\providecommand{\bibinfo}[2]{#2}
\providecommand{\BIBentrySTDinterwordspacing}{\spaceskip=0pt\relax}
\providecommand{\BIBentryALTinterwordstretchfactor}{4}
\providecommand{\BIBentryALTinterwordspacing}{\spaceskip=\fontdimen2\font plus
\BIBentryALTinterwordstretchfactor\fontdimen3\font minus
  \fontdimen4\font\relax}
\providecommand{\BIBforeignlanguage}[2]{{%
\expandafter\ifx\csname l@#1\endcsname\relax
\typeout{** WARNING: IEEEtran.bst: No hyphenation pattern has been}%
\typeout{** loaded for the language `#1'. Using the pattern for}%
\typeout{** the default language instead.}%
\else
\language=\csname l@#1\endcsname
\fi
#2}}
\providecommand{\BIBdecl}{\relax}
\BIBdecl

\bibitem{DIGITIMES}
DIGITIMES, ``Trends in the global ic design service market,'' \emph{online
  http://www.digitimes.com/news/a20120313RS400.html?chid=2}, vol. 2013, 2013.

\bibitem{6926108}
U.~Guin, D.~Forte, and M.~Tehranipoor, ``{Anti-counterfeit Techniques: From
  Design to Resign},'' in \emph{14th Int. Workshop on Microprocessor Test and
  Verification}, Dec 2013, pp. 89--94.

\bibitem{threatszamiri}
K.~Zamiri~Azar, H.~Mardani~Kamali, H.~Homayoun, and A.~Sasan, ``{Threats on
  Logic Locking: A Decade Later},'' in \emph{Proceedings of the 2019 on Great
  Lakes Symposium on VLSI}.\hskip 1em plus 0.5em minus 0.4em\relax New York,
  NY, USA: ACM, 2019, pp. 471--476.

\bibitem{8203496}
M.~Yasin and O.~Sinanoglu, ``{Evolution of logic locking},'' in \emph{IFIP/IEEE
  Int. Conference on Very Large Scale Integration (VLSI-SoC)}, Oct 2017, pp.
  1--6.

\bibitem{Keshavarz2018}
S.~Keshavarz, C.~Yu, S.~Ghandali, X.~Xu, and D.~Holcomb, ``{Survey on
  Applications of Formal Methods in Reverse Engineering and Intellectual
  Property Protection},'' \emph{Journal of Hardware and Systems Security},
  vol.~2, no.~3, Sep 2018.

\bibitem{srcRoshanisefat}
S.~Roshanisefat, H.~Mardani~Kamali, and A.~Sasan, ``{SRCLock: SAT-Resistant
  Cyclic Logic Locking for Protecting the Hardware},'' in \emph{Proceedings of
  the 2018 on Great Lakes Symposium on VLSI}.\hskip 1em plus 0.5em minus
  0.4em\relax ACM, 2018, pp. 153--158.

\bibitem{comakimia}
K.~Zamiri~Azar, F.~Farahmand, H.~Mardani~Kamali, S.~Roshanisefat, H.~Homayoun,
  W.~Diehl, K.~Gaj, and A.~Sasan, ``{COMA}: Communication and obfuscation
  management architecture,'' in \emph{22nd International Symposium on Research
  in Attacks, Intrusions and Defenses ({RAID} 2019)}.\hskip 1em plus 0.5em
  minus 0.4em\relax {USENIX} Association, Sep. 2019, pp. 181--195.

\bibitem{7140252}
P.~Subramanyan, S.~Ray, and S.~Malik, ``{Evaluating the Security of Logic
  Encryption Algorithms},'' in \emph{IEEE Int. Symp. on Hardware Oriented
  Security and Trust (HOST)}, May 2015, pp. 137--143.

\bibitem{el2015integrated}
M.~El~Massad, S.~Garg, and M.~V. Tripunitara, ``{Integrated Circuit (IC)
  Decamouflaging: Reverse Engineering Camouflaged ICs within Minutes},'' in
  \emph{NDSS}, 2015.

\bibitem{8474189}
S.~{Roshanisefat}, H.~K. {Thirumala}, K.~{Gaj}, H.~{Homayoun}, and A.~{Sasan},
  ``{Benchmarking the Capabilities and Limitations of SAT Solvers in Defeating
  Obfuscation Schemes},'' in \emph{2018 IEEE 24th International Symposium on
  On-Line Testing And Robust System Design (IOLTS)}, July 2018, pp. 275--280.

\bibitem{7858346}
M.~Yasin, B.~Mazumdar, O.~Sinanoglu, and J.~Rajendran, ``{Security Analysis of
  Anti-SAT},'' in \emph{22nd Asia and South Pacific Design Automation Conf.},
  Jan 2017.

\bibitem{Shamsi:2017:COC:3060403.3060458}
K.~Shamsi, M.~Li, T.~Meade, Z.~Zhao, D.~Z. Pan, and Y.~Jin, ``{Cyclic
  Obfuscation for Creating SAT-Unresolvable Circuits},'' in \emph{Proc. of the
  on Great Lakes Symposium on VLSI 2017}.\hskip 1em plus 0.5em minus
  0.4em\relax ACM, 2017, pp. 173--178.

\bibitem{8203759}
H.~Zhou, R.~Jiang, and S.~Kong, ``{CycSAT: SAT-based Attack on Cyclic Logic
  Encryptions},'' in \emph{IEEE Int. Conf. on Computer-Aided Design}, 2017, pp.
  49--56.

\bibitem{rajendran2013security}
J.~Rajendran, M.~Sam, O.~Sinanoglu, and R.~Karri, ``{Security Analysis of
  Integrated Circuit Camouflaging},'' in \emph{Proceedings of the 2013 ACM
  SIGSAC Conf. on Computer \& Communications Security}.\hskip 1em plus 0.5em
  minus 0.4em\relax ACM, 2013, pp. 709--720.

\bibitem{6881480}
R.~P. Cocchi, J.~P. Baukus, L.~W. Chow, and B.~J. Wang, ``{Circuit Camouflage
  Integration for Hardware IP Protection},'' in \emph{51st IEEE Design
  Automation Conf. (DAC)}, June 2014, pp. 1--5.

\bibitem{7495587}
B.~Erbagci, C.~Erbagci, N.~E.~C. Akkaya, and K.~Mai, ``A secure camouflaged
  threshold voltage defined logic family,'' in \emph{2016 IEEE International
  Symposium on Hardware Oriented Security and Trust (HOST)}, May 2016, pp.
  229--235.

\bibitem{li2017provably}
M.~Li, K.~Shamsi, T.~Meade, Z.~Zhao, B.~Yu, Y.~Jin, and D.~Z. Pan, ``{Provably
  Secure Camouflaging Strategy for IC Protection},'' \emph{IEEE Transactions on
  Computer-Aided Design of Integrated Circuits and Systems}, 2017.

\bibitem{Yasin_sfll}
M.~Yasin, A.~Sengupta, M.~T. Nabeel, M.~Ashraf, J.~J. Rajendran, and
  O.~Sinanoglu, ``{Provably-Secure Logic Locking: From Theory To Practice},''
  in \emph{Proceedings of the 2017 ACM SIGSAC Conf. on Computer and Comm.
  Security}, 2017, pp. 1601--1618.

\bibitem{8429401}
H.~Mardani~Kamali, K.~Zamiri~Azar, K.~Gaj, H.~Homayoun, and A.~Sasan,
  ``{LUT-Lock: A Novel LUT-Based Logic Obfuscation for FPGA-Bitstream and
  ASIC-Hardware Protection},'' in \emph{IEEE Computer Society Symp. on VLSI},
  July 2018, pp. 405--410.

\bibitem{smtazar}
K.~Zamiri~Azar, H.~Mardani~Kamali, H.~Homayoun, and A.~Sasan, ``{SMT Attack:
  Next Generation Attack on Obfuscated Circuits with Capabilities and
  Performance Beyond the SAT Attacks},'' \emph{IACR Transactions on
  Cryptographic Hardware and Embedded Systems}, vol. 2019, no.~1, pp. 97--122,
  Nov. 2018.

\bibitem{7495588}
M.~Yasin, B.~Mazumdar, J.~J.~V. Rajendran, and O.~Sinanoglu, ``{SARLock: SAT
  Attack Resistant Logic Locking},'' in \emph{2016 IEEE Int. Symp. on Hardware
  Oriented Security and Trust (HOST)}, May 2016, pp. 236--241.

\bibitem{xie2016mitigating}
Y.~Xie and A.~Srivastava, ``{Mitigating SAT Attack on Logic Locking},'' in
  \emph{Int. Conf. on Cryptographic Hardware and Embedded Systems}.\hskip 1em
  plus 0.5em minus 0.4em\relax Springer, 2016.

\bibitem{7951805}
K.~Shamsi, M.~Li, T.~Meade, Z.~Zhao, D.~Z. Pan, and Y.~Jin, ``{AppSAT:
  Approximately Deobfuscating Integrated Circuits},'' in \emph{IEEE Int'l Symp.
  on Hardware Oriented Security and Trust (HOST)}, 2017, pp. 95--100.

\bibitem{bypass}
X.~Xu, B.~Shakya, M.~M. Tehranipoor, and D.~Forte, ``{Novel Bypass Attack and
  BDD-based Tradeoff Analysis Against All Known Logic Locking Attacks},'' in
  \emph{Cryptographic Hardware and Embedded Systems (CHES)}.\hskip 1em plus
  0.5em minus 0.4em\relax Springer, 2017.

\bibitem{fall_attack}
D.~Sirone and P.~Subramanyan, ``{Functional Analysis Attacks on Logic
  Locking},'' in \emph{Design, Automation Test in Europe Conference Exhibition
  (DATE)}, 2019.

\bibitem{crosslock}
K.~Shamsi, M.~Li, D.~Z. Pan, and Y.~Jin, ``{Cross-Lock: Dense Layout-Level
  Interconnect Locking Using Cross-bar Architectures},'' in \emph{Proceedings
  of the 2018 on Great Lakes Symposium on VLSI}.\hskip 1em plus 0.5em minus
  0.4em\relax ACM, 2018, pp. 147--152.

\bibitem{Kamali:2019:FHD:3316781.3317831}
H.~Mardani~Kamali, K.~Zamiri~Azar, H.~Homayoun, and A.~Sasan, ``{Full-Lock:
  Hard Distributions of SAT Instances for Obfuscating Circuits Using Fully
  Configurable Logic and Routing Blocks},'' in \emph{Proceedings of the 56th
  Annual Design Automation Conference 2019}.\hskip 1em plus 0.5em minus
  0.4em\relax ACM, 2019, pp. 89:1--89:6.

\bibitem{kolhe2019security}
{G. Kolhe, H. Mardani Kamali, M. Naicker, T. D. Sheaves, H. Mahmoodi, S. M. P.
  Dinakarrao, H. Homayoun, S. Rafatirad, and A. Sasan}, ``{Security and
  Complexity Analysis of LUT-based Obfuscation: From Blueprint to Reality},''
  in \emph{Proceeding of the International Conference on Computer-Aided Design
  (ICCAD)}, 2019, pp. 1--8.

\bibitem{kolhecustom}
G.~Kolhe, S.~M. PD, S.~Rafatirad, H.~Mahmoodi, A.~Sasan, and H.~Homayoun, ``{On
  Custom LUT-Based Obfuscation},'' in \emph{Proceedings of the 2019 on Great
  Lakes Symposium on VLSI}, ser. GLSVLSI ’19.\hskip 1em plus 0.5em minus
  0.4em\relax New York, NY, USA: Association for Computing Machinery, 2019, p.
  477–482.

\bibitem{Tseitin1983}
G.~S. Tseitin, \emph{On the Complexity of Derivation in Propositional
  Calculus}.\hskip 1em plus 0.5em minus 0.4em\relax Springer Berlin Heidelberg,
  1983, pp. 466--483.

\bibitem{delaylocking}
Y.~Xie and A.~Srivastava, ``{Delay Locking: Security Enhancement of Logic
  Locking Against IC Counterfeiting and Overproduction},'' in \emph{Proceedings
  of the 54th Annual Design Automation Conference 2017}.\hskip 1em plus 0.5em
  minus 0.4em\relax ACM, 2017, pp. 9:1--9:6.

\bibitem{TimingSAT}
A.~Chakraborty, Y.~Liu, and A.~Srivastava, ``{TimingSAT: Timing Profile
  Embedded SAT Attack},'' in \emph{Proceedings of the Int. Conference on
  Computer-Aided Design}.\hskip 1em plus 0.5em minus 0.4em\relax ACM, 2018, pp.
  1--6.

\bibitem{Chen:2018:ESA:3217208.3190853}
Y.-C. Chen, ``{Enhancements to SAT Attack: Speedup and Breaking Cyclic Logic
  Encryption},'' \emph{ACM Trans. Des. Autom. Electron. Syst.}, vol.~23, no.~4,
  May 2018.

\bibitem{besat}
Y.~Shen, Y.~Li, A.~Rezaei, S.~Kong, D.~Dlott, and H.~Zhou, ``{BeSAT: Behavioral
  SAT-based Attack on Cyclic Logic Encryption},'' in \emph{Proceedings of the
  24th Asia and South Pacific Design Automation Conference}.\hskip 1em plus
  0.5em minus 0.4em\relax ACM, 2019, pp. 657--662.

\bibitem{ZhouCycles}
A.~Rezaei, Y.~Shen, S.~Kong, J.~Gu, and H.~Zhou, ``Cyclic locking and
  memristor-based obfuscation against cycsat and inside foundry attacks,'' in
  \emph{Design, Automation Test in Europe Conference Exhibition (DATE)}, March
  2018, pp. 85--90.

\bibitem{CycSATunresolvable}
A.~Rezaei, Y.~Li, Y.~Shen, S.~Kong, and H.~Zhou, ``{CycSAT-unresolvable Cyclic
  Logic Encryption Using Unreachable States},'' in \emph{Proceedings of the
  24th Asia and South Pacific Design Automation Conference}.\hskip 1em plus
  0.5em minus 0.4em\relax ACM, 2019, pp. 358--363.

\bibitem{7406959}
J.~H. Chen, Y.~C. Chen, W.~C. Weng, C.~Y. Huang, and C.~Y. Wang, ``Synthesis
  and verification of cyclic combinational circuits,'' in \emph{IEEE Int'l
  System-on-Chip Conf. (SOCC)}, 2015, pp. 257--262.

\bibitem{1466160}
V.~Agarwal, N.~Kankani, R.~Rao, S.~Bhardwaj, and J.~Wang, ``An efficient
  combinationality check technique for the synthesis of cyclic combinational
  circuits,'' in \emph{Proc. of the ASP-DAC}, 2005, pp. 212--215.

\bibitem{1218927}
M.~D. Riedel and J.~Bruck, ``The synthesis of cyclic combinational circuits,''
  in \emph{Proc. 2003. Design Automation Conf.}, 2003, pp. 163--168.

\bibitem{Rivest:1977:NFM:1310165.1310794}
R.~L. Rivest, ``{The Necessity of Feedback in Minimal Monotone Combinational
  Circuits},'' \emph{IEEE TC}, vol.~26, no.~6, pp. 606--607, 1977.

\bibitem{hawick2008enumerating}
K.~A. Hawick and H.~A. James, ``{Enumerating Circuits and Loops in Graphs with
  Self-Arcs and Multiple-Arcs.}'' in \emph{FCS}, 2008, pp. 14--20.

\bibitem{10.1007/978-3-319-08867-9_49}
B.~Dutertre, ``Yices 2.2,'' in \emph{Computer Aided Verification}.\hskip 1em
  plus 0.5em minus 0.4em\relax Springer, 2014, pp. 737--744.

\bibitem{irataVakil}
A.~Vakil, H.~Homayoun, and A.~Sasan, ``{IR-ATA: IR Annotated Timing Analysis, a
  Flow for Closing the Loop Between PDN Design, IR Analysis \& Timing
  Closure},'' in \emph{Proceedings of the 24th Asia and South Pacific Design
  Automation Conference}.\hskip 1em plus 0.5em minus 0.4em\relax ACM, 2019, pp.
  152--159.

\end{thebibliography}

\vspace{-35pt}
\begin{IEEEbiography}[{\includegraphics[width=1in,height=1.28in,clip,keepaspectratio]{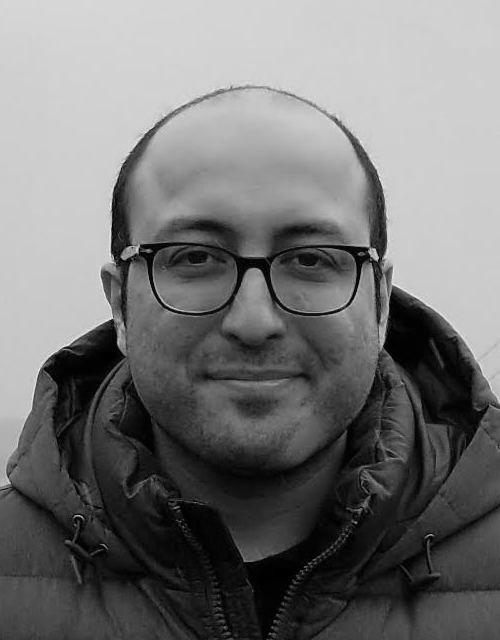}}]{Shervin~Roshanisefat}
received his B.Sc. degree in computer engineering from Azad University in Qazvin, Iran. He received his M.Sc. degree in computer engineering from the University of Tehran, Iran in 2015. He has worked earlier on telecommunication devices for SRD in Iran. He is currently a Ph.D. candidate at the Electrical and Computer Engineering department at George Mason University. His research interests are in the areas of approximate computing, low power design, hardware-level functional safety, and security.
\end{IEEEbiography}
\vspace{-30pt}
\begin{IEEEbiography}[{\includegraphics[width=1in,height=1.28in,clip,keepaspectratio]{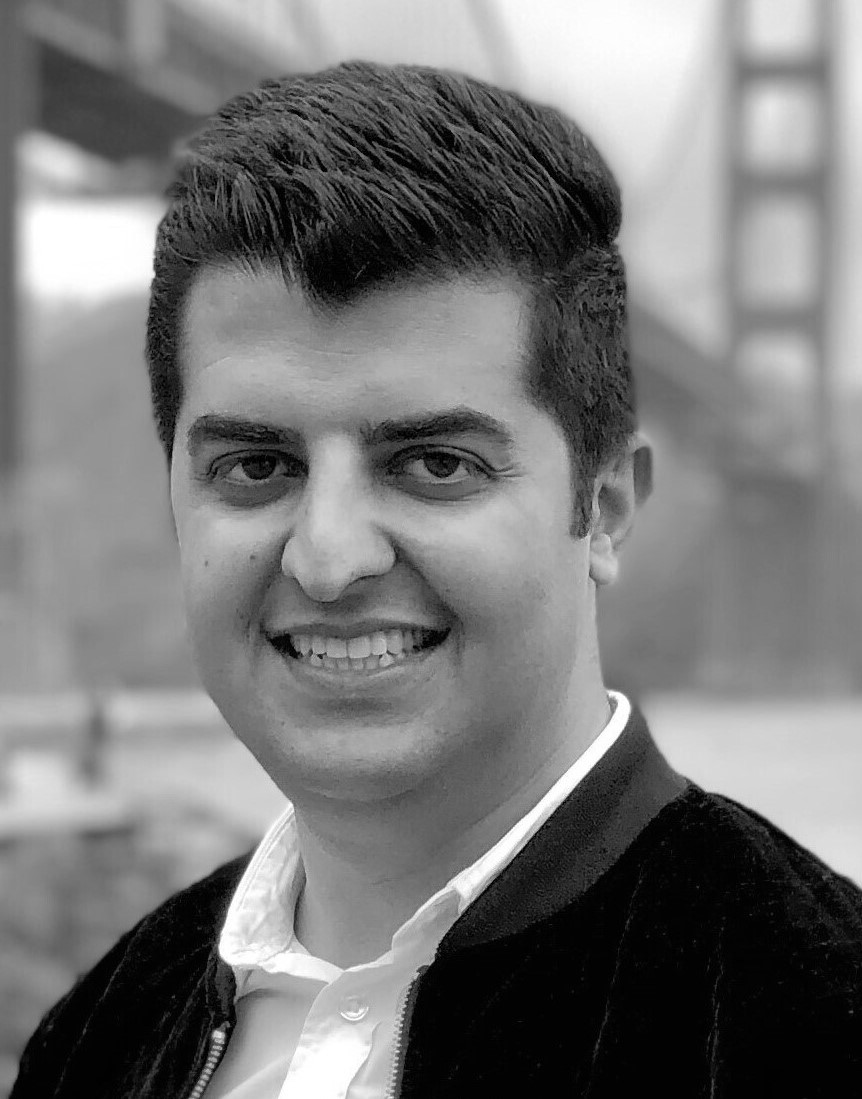}}]{Hadi Mardani Kamali}
received his B.Sc. degree in computer engineering from Khajeh Nasir University in Tehran, Iran in 2011, and his M.Sc. degree in computer engineering from Sharif University of Technology, Tehran, Iran in 2013. He is currently a Ph.D. student at the Electrical and Computer Engineering department at George Mason University. His research focuses on hardware security, hardware/software acceleration applications, and power management in on-chip communication.
\end{IEEEbiography}
\vspace{-35pt}
\begin{IEEEbiography}[{\includegraphics[width=1in,height=1.28in,clip,keepaspectratio]{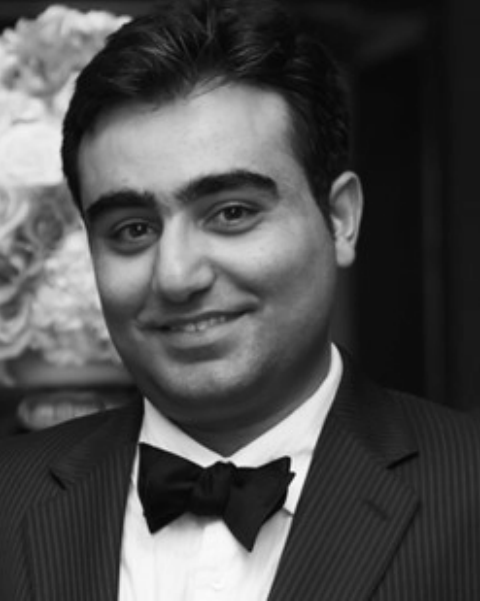}}]{Houman Homayoun}
received the B.Sc. degree in electrical engineering from Sharif University of Technology, Tehran, Iran, in 2003, the M.Sc. degree in computer engineering from the University of Victoria, Victoria, BC, Canada, in 2005, and the Ph.D. degree from the Department of Computer Science, University of California at Irvine, Irvine, CA, USA, in 2010. He is currently an Associate Professor at the Electrical and Computer Engineering Department, University of California Davis, CA, USA. Since 2017 he has been serving as an Associate Editor of IEEE Transactions on VLSI. He was the technical program co-chair of GLSVLSI 2018 and the general chair of the 2019 GLSVLSI conference.
\end{IEEEbiography}
\vspace{-30pt}
\begin{IEEEbiography}[{\includegraphics[width=1in,height=1.52in,clip,keepaspectratio]{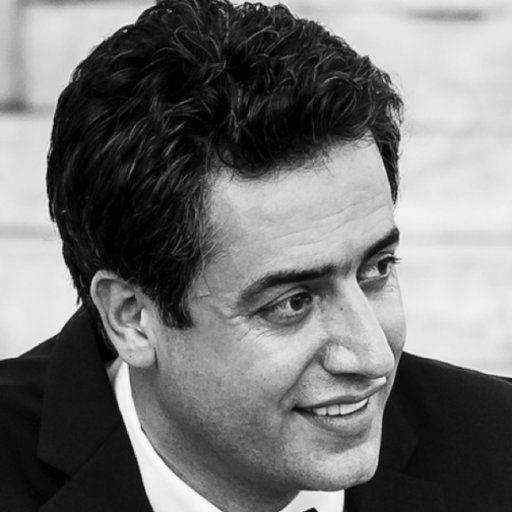}}]{Avesta Sasan}
received his B.Sc. in Computer Engineering from the University of California Irvine in 2005 with the highest honor (Summa Cum Laude). He then received his M.Sc. and his Ph.D. in Electrical and Computer Engineering from the University of California Irvine in 2006 and 2010 respectively. Dr. Sasan then worked at the industry in Broadcom and Qualcomm Co. till 2016. He joined George Mason University in 2016, where he is currently serving as an Associate Professor in the Department of Electrical and Computer Engineering.
\end{IEEEbiography}

\end{document}